%% file: TNGClusterSpiralMorphology.tex
\documentclass[fleqn,usenatbib,usenames,dvipsnames]{mnras}

\usepackage{newtxtext,newtxmath}
\usepackage[T1]{fontenc}
\usepackage{ae,aecompl}

\usepackage{graphicx}	% Including figure files
\usepackage{amsmath}	% Advanced maths commands
\usepackage{amssymb}	% Extra maths symbols
\usepackage{xcolor}
\usepackage{times}
\usepackage{soul}

\newcommand{\MSUN}{{\rm M}_{\sun}}

\title[The fate of IllustrisTNG cluster discs]{The fate of disc galaxies in IllustrisTNG clusters}

\author[G. D. Joshi et al.]{Gandhali D. Joshi,$^{1}$\thanks{E-mail: joshi@mpia.de}, 
Annalisa Pillepich$^{1}$, 
Dylan Nelson$^2$, 
Federico Marinacci$^3$,\newauthor
Volker Springel$^2$,
Vicente Rodriguez-Gomez$^4$,
Mark Vogelsberger$^5$,
and Lars Hernquist$^6$
 \\~\\
$^{1}$Max-Planck-Institut f{\"u}r Astronomie, K{\"o}nigstuhl 17, D-69117 Heidelberg, Germany\\
$^{2}$Max-Planck-Institut f{\"u}r Astrophysik, Karl-Schwarzschild-Str. 1, D-85748, Garching, Germany \\
$^{3}$ Department of Physics and Astronomy, University of Bologna, Via Gobetti 93/2, I-40129 Bologna, Italy \\
$^{4}$Instituto de Radioastronom{\'i}a y Astrof{\'i}sica, Universidad Nacional Aut{\'o}noma de M{\'e}xico, A.P. 72-3, 58089 Morelia, M{\'e}xico \\
$^{5}$Department of Physics, Massachusetts Institute of Technology, Cambridge, MA 02139, USA \\
$^{6}$Institute for Theory and Computation, Harvard-Smithsonian Center for Astrophysics, 60 Garden Street, Cambridge, MA 02138, USA \\
}

\date{Accepted XXX. Received YYY; in original form ZZZ}

\pubyear{2019}

\begin{document}
\label{firstpage}
\pagerange{\pageref{firstpage}--\pageref{lastpage}}
\maketitle

\begin{abstract}
We study the stellar morphological evolution of disc galaxies within clusters in the TNG50 and TNG100 runs from the IllustrisTNG simulation suite. We select satellites of masses $10^{9.7} \leq M_{*,z=0}/\MSUN \leq 10^{11.6}$ residing in clusters of masses $10^{14} \lesssim M_{\text{200c,z=0}}/\MSUN \leq 10^{14.6}$ at $z=0$ and that were discs at accretion according to a kinematic morphology indicator (the circularity fraction). These are traced from the time of accretion to $z=0$ and compared to a control sample of central galaxies mass-matched at accretion. Most cluster discs become non-discy by $z=0$, in stark contrast with the control discs, of which a significant fraction remains discy over the same timescales. Cluster discs become non-discy accompanied by gas removal and star formation quenching, loss of dark matter and little growth or a loss of stellar mass. In contrast, control discs transform while also losing gas mass and quenching, but growing significantly in dark matter and stellar mass. Most cluster satellites change morphologies on similar timescales regardless of stellar mass, in $\sim0.5-4$ Gyr after accretion. Cluster discs that experienced more numerous and closer pericentric passages show the largest change in morphology. Morphological change in all cases requires the presence of a gravitational perturbation to drive stellar orbits to non-discy configurations, along with gas removal/heating to prevent replenishment of the disc through continued star-formation. For cluster discs, the perturbation is impulsive tidal shocking at pericentres and not tidal stripping of outer disc stellar material, whereas for control discs, a combination of mergers and AGN feedback appears to be the key driving force behind morphological transformations.
\end{abstract}


%%%%%%%%%%%%%%%%%%%%%%%%%%%%%%%%%%%%%%%%%%%%%%%%%%

%%%%%%%%%%%%%%%%% BODY OF PAPER %%%%%%%%%%%%%%%%%%

\section{Introduction} \label{sec:intro}
Recent decades have seen a focused effort at understanding the evolution of galaxies in diverse environments. It is now well known that in dense environments, ranging from groups of a few galaxies to massive clusters containing thousands of galaxies, observed galaxy populations display distinct properties compared to isolated, field galaxy populations. Broadly speaking, group and cluster populations have higher proportions of red, quenched, elliptical galaxies, whereas field populations are dominated by blue, star-forming, spiral galaxies \citep[e.g.][]{Oemler74,Dressler80,Balogh04,Hogg04,Kauffmann04,Blanton05,Mehmet15}. Although several studies have explored the causes of the differences between these populations, key questions regarding the physical mechanisms driving these differences are yet to be answered.

Various environmental scenarios have been theoretically proposed as viable mechanisms that can affect galaxies in dense environments and contribute to their divergent evolutionary paths. Interactions with the cluster halo and the intra-cluster medium (ICM) can lead to tidal stripping, ram-pressure stripping and starvation. Tidal stripping by the stronger gravitational forces exerted by the cluster halo can not only remove matter from the galaxy, it can also produce distinct morphological features such as tidal tails \citep{Toomre72,Barnes92,Bournaud04}. Ram pressure stripping due to hydrodynamical pressure encountered by travelling through a background gaseous medium can remove even the more bound cold gas from the galaxy as well as produce features such as warped or bent discs or jellyfish-like gas tails \citep{Gunn72,Abadi99,Yun19}. On the other hand, starvation due to the cut-off of fresh gas accretion can gradually quench star formation in the galaxy without affecting its morphology significantly \citep{Larson80,Balogh00,Kawata08}. Interactions with nearby galaxies can also lead to harassment and mergers. High-speed encounters with nearby galaxies can make the stellar and dark matter components of the galaxy less bound and therefore more susceptible to stripping \citep{Moore96,Moore98}. Finally, mergers with other galaxies can not only spark bursts of star formation followed by a rapid decline in the star formation rate (SFR), they can also produce rapid changes in morphology \citep{Mihos96,Barnes96,Makino97,Angulo09}.

The efficiency of each of these mechanisms may depend on several factors including the mass of the cluster, the location of the galaxy within the cluster, the matter composition of the galaxy, the properties of the galaxy's orbit, and therefore collectively on the mass assembly history of the host. Understanding the evolution of different galactic properties can help in identifying which processes are dominant in such environments, at what times and over what timescales. So far, most of our theoretical understanding has been tested and evaluated through a landscape of controlled non-cosmological and/or N-body only numerical experiments.

Tidal interaction with a group or cluster is one key mechanism for transforming spiral galaxies into elliptical, or more generally non-discy, galaxies. Through N-body simulations of the collisionless dark matter (DM) and stellar components of a Milky Way (MW) mass disc galaxy in a controlled group-like environment, \citet{Villalobos12} found that structural changes only occur when the mean density enclosed by the galaxy's orbit is $\sim0.3-1$ times the central mean density of the galaxy. The impact of the group tidal field was found to be highly dependent on the galaxy disc inclination, with face-on galaxies and those with retrograde orbits retaining their disc structures longer than those on prograde orbits. At lower mass scales and also within controlled collisionless setups, \citet{Kazantzidis11} found that tidal interactions of equilibrium-model dwarf discs in MW-like hosts that lead to dwarf-spheroidal formation also depend on the orbital parameters of the galaxy, whereby small pericentric distances and short orbital times result in larger transformations. 

In addition to the group/cluster tidal field, harassment due to other member galaxies can also lead to morphological changes, as has been studied by several groups. \citet{Bekki11} showed with hydrodynamical experiments that spiral galaxies in idealized group environments can be transformed into S0s through multiple tidal encounters with other group members as well as the group potential, by dynamically heating the initial thin discs into thick discs, while also triggering bursts of star formation in the inner regions of the galaxies leading to bulge growth. The changes are strongly dependent on the galaxy's mass, the total group mass and the orbit of the galaxy. Comparable results were also found by \citet{Bialas15} with a combination of N-body simulations and Monte-Carlo methods aimed at mimicking cluster environments, who showed that galaxy harassment can affect the morphologies of a fraction of cluster galaxies, those that were accreted early and had undergone several galaxy-galaxy encounters. Additionally, such encounters appeared to be most efficient near the centres of clusters. Similarly, \cite{Lisker13} also showed with the Millennium II cosmological simulation equipped with a semi-analytic model that the formation of dwarf ellipticals in cluster environments is likely the result of several encounters over long timescales rather than a single recent transformation. On the other hand, for lower-mass galaxies, \citet{Smith10} concluded with controlled models of the tidal field of Virgo-like clusters that harassment may not be an important formation mechanism for dwarf ellipticals that were accreted recently.

Galaxy mergers are another channel through which disc stellar morphologies can be transformed into spheroidal ones; this appears to be the case both in the field and within group/cluster environments, but differently so for central and satellite galaxies. With a zoom-in cosmological hydrodynamical simulation of a $10^{13}\MSUN$ group, \citet{Feldmann11} studied the formation of a group of galaxies and found that elliptical galaxies in the group were formed through mergers that occurred well before accretion into the group, at $z>1$. They also found -- in the absence of feedback from super massive black holes -- that star formation quenching lagged behind the changes in morphology, starting when the galaxies approached the group's virial radius but occurring on short timescales once begun. While a number of studies \citep[e.g.][]{Martin18, Clauwens18} have analyzed the effects of minor and major mergers in driving the formation of massive spheroidal galaxies within modern large-scale cosmological simulations such as Horizon-AGN \citep{Dubois16} and EAGLE \citep{Crain15,Schaye15} and the original Illustris simulations \citep{RodriguezGomez17}, little emphasis has been placed specifically on satellite galaxies of massive groups and clusters. 

Cosmological hydrodynamical simulations and semi-analytical models (SAMs) typically allow us to directly follow the histories of individual galaxies, whereby we can determine the precise environmental conditions they encounter and connect these to their evolution and that of their host. They also provide a realistic and complete sampling of satellite accretion histories and orbits, as the latter naturally emerge from the hierarchical growth of structure that is self-consistently captured within such simulation setups. However, to quantify galaxy morphology trends in cosmological simulations requires high numerical resolution and detailed physics models which generate realistic populations of disc and elliptical galaxies. On the other hand, large simulation volumes or large numbers of ``zoom-in'' cluster simulations are needed to conduct robust statistical analyses and to sample rare, high-mass hosts. Both these requirements have significant computational costs and have therefore only been met in recent years. 

Notable examples of cosmological hydrodynamical simulations of galaxies with comprehensive physical models in group and cluster environments (i.e. with total host mass exceeding a few $10^{13}\MSUN$) include the aforementioned large-scale uniform volume simulations: e.g. Illustris \citep{Vogelsberger14, Genel14, Sijacki15}, TNG100 of the IllustrisTNG suite \citep{TNGSpringel18, TNGNelson18, TNGNaiman18, TNGPillepich18, TNGMarinacci18}, EAGLE, Horizon-AGN, Magneticum \citep{Hirschmann14,Dolag15,Dolag16}, Mufasa \citep{Dave16}, and Simba \citep{Dave19}. With kpc-like spatial resolution and stellar particle masses of about $10^{6-7}\MSUN$, these state-of the-art models have been more recently augmented on two fronts: (i) with larger samples of groups and more massive clusters up to $\sim10^{15}\MSUN$ at similar resolution, i.e. with the TNG300 box, the largest volume of the IllustrisTNG suite, and the suite of 24 (Hydrangea) plus 6 additional (C-EAGLE) zoom-in simulations of massive galaxy clusters \citep{Bahe17,Barnes17}, or (ii) with enhanced numerical resolution for Virgo-mass like hosts, specifically with the TNG50 run \citep[the highest-resolution implementation of the IllustrisTNG project,][]{TNG50Pillepich19, TNG50Nelson19} and the Romulus-C cluster zoom simulation \citep{Tremmel19}.

In this study, we leverage the combination of the TNG100 and TNG50 cosmological magneto-hydrodynamical simulations to study the morphological evolution of satellite galaxies in Virgo-like clusters: the goal is to quantify the role played by the environment, to understand how it can induce changes to the stellar structures and kinematics of satellite galaxies, and hence to identify what physical processes play a key role in this transformation. To this end, we select a population of galaxies that are found today within clusters of mass $M_{\text{200c}}\sim10^{14-14.6}\MSUN$\footnote{$M_{\text{200c}}$ is the mass enclosed by a radius within which the mean density is 200 times the critical density at the given redshift.} and that were discs just prior to their accretion on their final hosts, and follow their morphological evolution in detail. We are then able to compare their transformation in morphology with changes in their star formation activity and gas content. We contrast against a sample of similar field galaxies over the same timescales, thereby isolating the role of the cluster environment from secular (e.g. black-hole feedback) and other assembly processes (e.g. mergers) that may generally affect the stellar morphologies of galaxies regardless of their environment. We rely on a galaxy formation model that has been shown to return a population of quenched galaxies even at intermediate/high redshifts \citep{Weinberger18,Donnari19}, a good match of the shapes and widths of the red sequence and blue cloud in comparison to SDSS galaxies \citep{TNGNelson18}, an observationally-consistent spatial clustering of red vs. blue galaxies over tens of kpc to tens of Mpc distances \citep{TNGSpringel18}, and -- most interestingly -- an observationally-consistent separation of star-forming and quenched galaxies in a number of galaxy observables across cosmic epochs. These include stellar sizes \citep{Genel18} and the stellar morphologies of central or all galaxies with no distinction in environment \citep{TNGNelson18,RodriguezGomez19,Tacchella19}.

The paper is organized as follows: in Section \ref{sec:methods}, we provide details about the simulation and measurements used in the rest of the paper and in Section \ref{sec:samples}, we define our galaxy sample selection. Section \ref{sec:clusterDiscs} examines the degree of change in their stellar morphology and compares it to change in other galaxy properties. In Section \ref{sec:rates}, we study the timescales over which morphological transformation occurs. We explore the dependence of the degree and timescale of morphological change on various environmental and intrinsic factors in Section \ref{sec:keyFactors}. Finally, we discuss our results in Section \ref{sec:discussion} and lay out our conclusions in Section \ref{sec:conclusion}.

% ------------------------------------------------------------------------------------------------------------

\begin{figure*}
	\includegraphics[width=\linewidth]{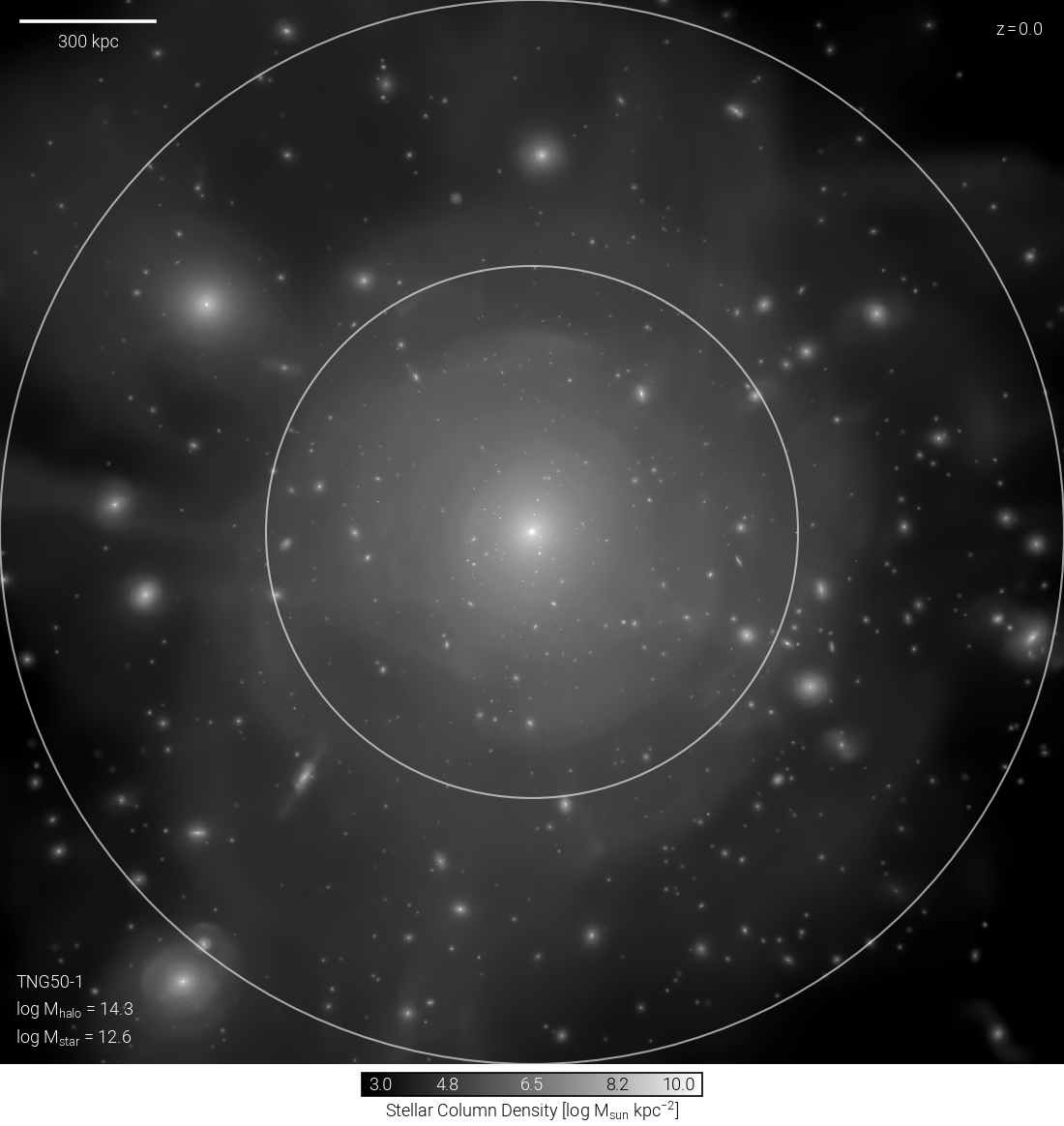}
	\caption{Stellar mass density in a random projection of the most massive halo in TNG50 at $z=0$. Circles denote half and one times the virial radius of the cluster. In TNG50, the stellar particle mass is $8.5\times 10^4 \MSUN$, making this the highest resolution simulation of a Virgo-like cluster.} \label{fig:tng50_halo0}
\end{figure*}

\begin{figure*}
	\includegraphics[width=\linewidth]{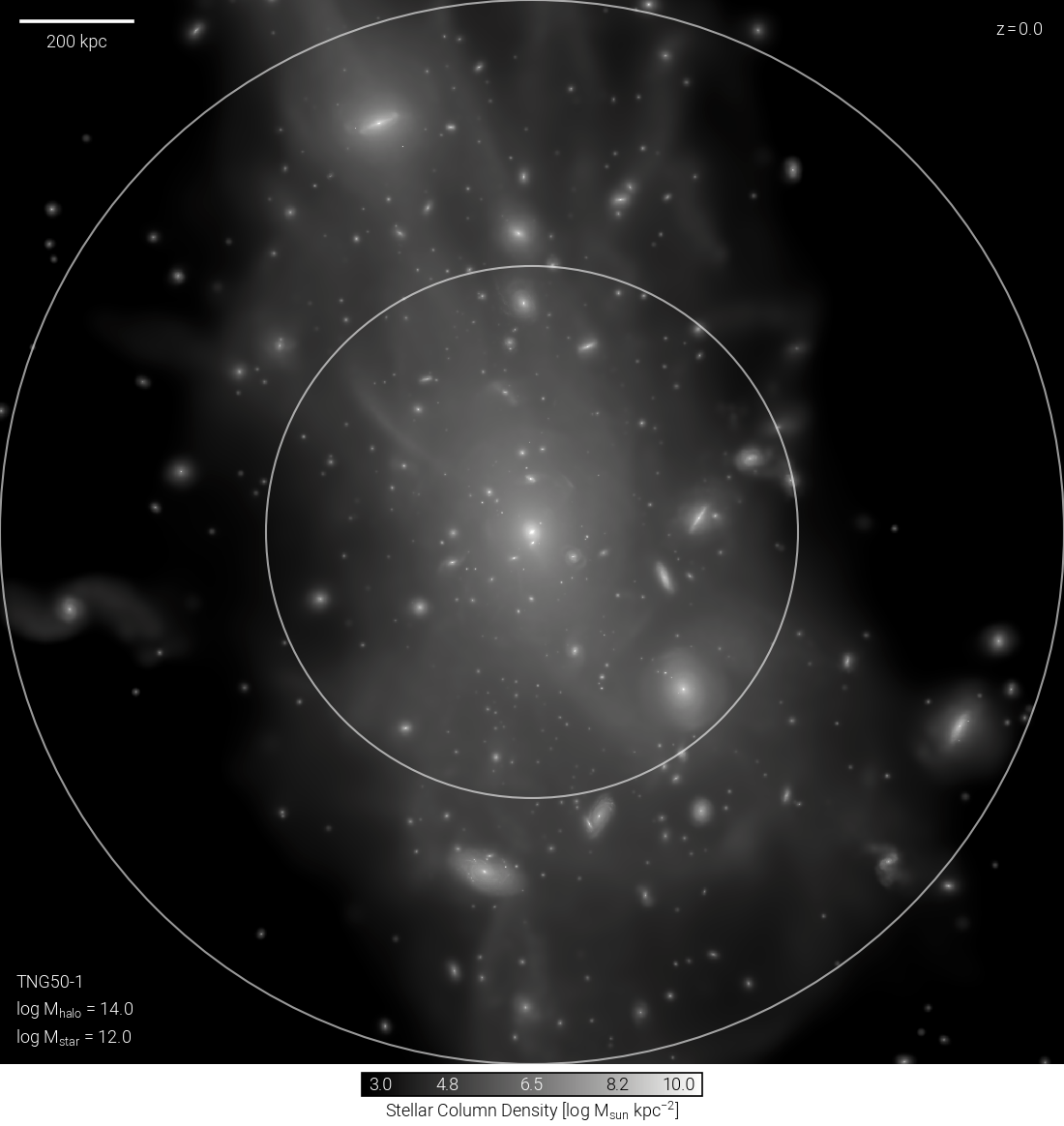}
	\caption{Stellar mass density in a random projection of the second most massive halo in TNG50 at $z=0$. Annotations as in Figure~\ref{fig:tng50_halo0}.} \label{fig:tng50_halo1}
\end{figure*}

\section{Methods} \label{sec:methods}

\subsection{Simulations}
We use the results from IllustrisTNG which is a suite of cosmological simulations covering three volumes of (35 $h^{-1}$Mpc)$^3$, (75 $h^{-1}$Mpc)$^3$ and (205 $h^{-1}$Mpc)$^3$ (TNG50, TNG100 and TNG300 respectively); specifically, we use the results of TNG100 \citep{TNGNelson18,TNGNaiman18,TNGPillepich18,TNGSpringel18,TNGMarinacci18,TNGDRNelson19} and TNG50 \citep{TNG50Pillepich19,TNG50Nelson19}. The simulations are run using the moving-mesh code \textsc{arepo} \citep{Springel10}, which adopts a tree-particle-mesh algorithm to solve Poisson's equations for gravity, and a finite volume scheme on a moving, unstructured mesh, generated through Voronoi tessellation of the simulation volume, to solve the equations of ideal MHD. The underlying physics models are described in \citet{TNGMethodsWeinberger17,TNGMethodsPillepich18}. The IllustrisTNG model includes prescriptions for gas heating by an evolving UV background, radiative gas cooling including primordial and metal-line cooling, star formation in dense regions and stellar evolution and black-hole seeding and growth through accretion and mergers. It also models feedback from supernovae in the form of galactic winds, and from active galactic nuclei (AGN) in the form of thermal energy injection during the high-accretion mode and kinetic energy during the low-accretion mode. Metal enrichment from nine different elements is tracked separately for SNe-Ia, SNe-II and asymptotic giant branch stars. The IllustrisTNG suite also includes lower resolution runs for each of the three volumes as well as dark matter-only counterparts for each hydrodynamical run. All simulations employ a flat $\Lambda$CDM cosmology using cosmological parameters from \citet{Planck15} with $h=0.6774$, $\Omega_{\text{m}}=0.3089$, $\Omega_{\Lambda}=0.6911$, $\Omega_{\text{b}}=0.0486$, $\sigma_{8}=0.8159$ and $n_{s}=0.9667$. Each run consists of 100 snapshots from $z\sim 20$ to $z=0$.

In this study, we make use of the results of the TNG50 and TNG100 simulations. The TNG50 (TNG100) run simulates a cubic volume of comoving side length 35 (75) $h^{-1}$Mpc ($\approx$51.7 (110.7) Mpc), initially with $2\times2160^{3}$ ($2\times1820^{3}$) resolution elements in the form of equal numbers of dark matter particles and gas cells. This results in DM particles of constant mass $m_{\text{DM}}=4.5\times10^{5}\MSUN$ ($7.5\times10^{6}\MSUN$) and gas cells or stellar particles with an average mass of $m_{\text{bary}}=8.5\times10^{4}\MSUN$ ($1.4\times10^{6}\MSUN$). Gravitational softening lengths for the stellar and DM particles are 288 (738) pc at $z=0$, comoving until $z=1$ and physical at lower redshifts. The gas softening lengths are adaptive, set to 2.5 times the comoving gas cell radius, with a minimum value of 73.8 (184.5) pc \citep[see e.g. figure 1 of][]{TNG50Pillepich19}.

Galaxies are identified within the snapshots in a two-step process. First, dark matter haloes are detected using a Friends-of Friends (FOF) algorithm \citep{Davis85} with a linking length of $b=0.2$ times the mean inter-particle spacing. Note that the FOF algorithm is only run on the dark matter particles and baryonic particles are assigned to the FOFs containing the dark matter particle nearest to them. Galaxies are then identified within each FOF group using \textsc{subfind} \citep{Springel01,Dolag09}, which detects gravitationally bound substructures within the FOF groups (using DM and baryonic particles). Several properties of the haloes and subhaloes are provided by the two algorithms. After galaxies were identified at each snapshot, merger trees were generated by tracking their baryonic content using the \textsc{sublink} \citep{RodriguezGomez15} algorithm and we use these `SubLink\_gal' trees in the present work.

\subsection{General galaxy properties}

In this analysis, we consider several properties of the galaxies as identified by \textsc{subfind} and through post-processing of the simulation results. A galaxy is defined as any subhalo with a nonzero total stellar mass and nonzero dark matter mass at $z=0$. The \textsc{subfind} occasionally identifies subhaloes that are unlikely to be of cosmological origin i.e. not formed through cosmological structure formation. Instead, these objects may form through fragmentation of baryonic material within galaxies already formed. We make use of the `SubhaloFlag' defined in \citet{TNGDRNelson19} to remove such clumps. Briefly, the procedure identifies subhaloes that were satellites at the time of formation, formed within $0.1r_\text{vir}$ of their host halo and had a DM fraction i.e. total bound DM to total bound mass of $<0.8$ at formation and flags them as unlikely to be real galaxies. The definitions of the key properties are provided below.

\paragraph*{Positions and velocities:}
The position of a galaxy is given by the position of its minimum potential particle, of any type, and its velocity is the mass weighted sum of \emph{all} particles belonging to the galaxy, regardless of type. In the case of FOF groups, the position of the FOF group is identical to that of its central galaxy.

\paragraph*{Mass and SFR:}
For this study, we measure all halo properties within the virial radius $r_{200c}$ of the FOF group, encompassing an average density that is 200 times the critical density of the Universe (note that this includes \emph{all} particles contained within this radius). Galaxy properties are measured within twice the stellar half mass radius, $r_{\text{*,1/2}}$ (note that in this case, only gravitationally bound particles are considered). Therefore, for the remainder of this paper, the radius $r$, mass $M$, and the mass of any component, i.e. $M_{\text{*}}$, $M_{\text{gas}}$, $M_{\text{DM}}$, of a galaxy refer to the stellar half mass radius and the mass contained within twice the stellar half mass radius, unless specified otherwise. The SFR of the galaxy is also measured within $2\,r_{\text{*,1/2}}$, and is the sum of the instantaneous SFRs of all the gravitationally bound gas cells contained within it.

\subsection{Galaxy stellar morphology} \label{sec:methodsMorph}
We quantify the stellar morphologies of our galaxies using the circularity fraction, $\mathit{f}_{\epsilon}$, which measures the mass fraction of stellar particles that are part of the galaxy disc, thereby providing a kinematic measure of its morphology. Note that morphologies are only calculated for galaxies with $M_{*}>10^{8}\MSUN$ for TNG50 and $M_{*}>10^{9}\MSUN$ for TNG100. We consider these measurements to be robust for galaxies with $M_{*}>3\times10^{8}\MSUN$ and $M_{*}>5\times10^{9}\MSUN$ for the two runs respectively, i.e. when calculated using at least $\approx$ 3500 particles; measurements at lower masses are included for completeness, but should be interpreted with this caveat.

We follow the methodology of \citet{Genel15} to characterize the kinematic stellar morphology of the galaxies. The circularity parameter $\epsilon$ is defined following \citet{Marinacci14}, as 
\begin{equation}
    \epsilon = \frac{J_{z}}{J_{\text{max}}(E)}
\end{equation}
where $J_{z}$ is the specific angular momentum of a given stellar particle parallel to the disc of the galaxy, and $J_{\text{max}}(E)$ is its maximum possible angular momentum, where $E$ is its specific binding energy. For these calculations, while the galaxy's centre is taken to be the position of its minimum potential particle (of any type), the reference velocity is taken to be the mass weighted average velocity of the stellar particles. We only consider particles within $2\,r_{\text{*,1/2}}$ for these calculations. The disc orientation is then defined by the total angular momentum of the stellar particles. In order to calculate $J_{\text{max}}(E)$, we follow the procedure of \citet{Genel15} by arranging the particles according to their specific binding energy $E$ and taking the maximum $J_{z}$ of the 50 particles before and 50 after the particle in question.

The `disciness' of the galaxy is measured by $\mathit{f}_{\epsilon}$, the mass fraction of stellar particles that have a circularity $\epsilon>0.7$. The threshold of $\epsilon>0.7$ has been shown to reliably select orbits confined to a disc \citep[e.g.][]{Aumer13}. We consider galaxies to be discy if $\mathit{f}_{\epsilon}>0.4$ and non-discy if $\mathit{f}_{\epsilon}<0.4$, where the critical fraction was selected through visual inspection of edge-on images of our galaxy sample. Note that the highest value of $\mathit{f}_{\epsilon}$ we measure for any galaxy at any time is 0.78.

As shown in later sections, this kinematic morphology indicator does not always match visual classifications of morphology in maps of the galaxy's photometry. While high values of $\mathit{f}_{\epsilon}$ ($\gtrsim0.5$) do always correspond to galaxies that appear as thin discs, often with clear spiral arms, galaxies with $\mathit{f}_{\epsilon}$ just below the critical value of $0.4$ still appear discy. Even lower values of $\mathit{f}_{\epsilon}$ can manifest as spheroidal stellar distributions, puffed-up discs, discs with tidal tails or perturbed features or galaxies with prolate shapes. This lack of correlation will be explored in a future study.

\subsection{Timescale for morphological transformation} \label{sec:methodsTimescales}
In order to understand the rate at which galaxies transform from discy to non-discy morphologies, we define a timescale to become non-discy, $\Delta t_{\text{nondisc,last}}$ as the time between accretion and the \emph{last} time the galaxy's morphology transitions from discy to non-discy, i.e. the last time $\mathit{f}_{\epsilon}$ changes from $>0.4$ to $\leq0.4$. The accretion time is defined in the following section. As shown in later sections, the morphological transformation is not necessarily a monotonous decline in $\mathit{f_{\epsilon}}$; using the last time the galaxy becomes non-discy allows us to capture the entire process of the morphological change after which the galaxy becomes permanently non-discy. We only calculate this timescale for galaxies that are discs at accretion (i.e. $\mathit{f_{\epsilon}}(\text{acc.})>0.4$) and non-discs at redshift zero (i.e. $\mathit{f_{\epsilon}}(z=0)\leq0.4$). Note however that this timescale only pinpoints the time at which the galaxy's circularity fraction drops below our chosen threshold and can be thought of as the beginning of the last phase of morphological transformation. Alternatively, we could define $\Delta t_{\text{nondisc,last}}$ as the last time the circularity fraction transitions from $>0.2$ to $\leq0.2$, whereby the galaxy has less than half the required disc mass fraction to be truly considered non-discy. We discuss the impacts of using this alternative definition in later sections.

% ------------------------------------------------------------------------------------------------------------

\begin{table}
    \caption{Properties of the host clusters for the TNG50 and TNG100 runs and their galaxy populations. We provide the clusters' virial mass and virial radius at $z=0$ together with some demographics of their satellite populations: the total number of \emph{luminous} satellite galaxies, i.e. with $M_{\text{*}}>0$, found within the virial radius $r_{\text{200c}}$ at $z=0$ (indicated as `lum'); the total number of satellite galaxies found within the virial radius $r_{\text{200c}}$ at $z=0$ with a minimum stellar mass as indicated; among the latter, the number of satellites that were discs at the time of accretion onto their $z=0$ host; and the number of those discs that remain discs at $z=0$.} \label{tab:hostProps}
    \centering
    \begin{tabular}{l|c|c|c|c|c|c}
    \hline
    & & & \multicolumn{4}{c}{\# cluster satellites} \\
    Label & $M_{\text{200c}}$  & $r_{\text{200c}}$ & lum. & tot & disc & discs \\
     & $[10^{14}\MSUN]$ & [Mpc] & & & at acc. & at $z=0$ \\
    \hline
    \multicolumn{7}{c}{TNG50 ($\log{M_{\text{*,z=0}}}=8.5-11.6$)} \\
    \hline
    \input{host_props_TNG50.tex}
    \hline
    \multicolumn{7}{c}{TNG100 ($\log{M_{\text{*,z=0}}}=9.7-12.6$)} \\
    \hline
    \input{host_props_TNG100.tex}
    \hline
    \end{tabular}
\end{table}

\begin{table}
    \caption{Sample sizes of cluster satellites and control galaxies and their respective disc subsamples at accretion, and the number of those disc galaxies that remain discs at $z=0$, for the TNG50 and TNG100 runs. Satellites are selected to be within the virial radius $r_{\text{200c}}$ of clusters of $M_{\text{200c,z=0}}\gtrsim 10^{14}\MSUN$ at $z=0$ (see Table \ref{tab:hostProps}).} \label{tab:sample}
    \centering
    \begin{tabular}{l|c|c|c|c|c|c}
    \hline
    & \multicolumn{3}{c}{\# Cluster satellites} & \multicolumn{3}{c}{\# Control galaxies} \\
    $\log{M_{\text{acc}}}$ & tot & discs & discs & tot & discs & discs \\
    & & at acc. & at $z=0$ & & at acc. & at $z=0$ \\
    \hline
    \multicolumn{7}{c}{TNG50} \\
    \hline
    \input{sample_TNG50.tex}
    \hline
    \multicolumn{7}{c}{TNG100} \\
    \hline
    \input{sample_TNG100.tex}
    \hline
    \end{tabular}
\end{table}

\section{Sample selection in TNG100 and TNG50} \label{sec:samples}

\subsection{Cluster parent sample} \label{sec:sampleZ0}
The goal of this study is to understand the effects that the cluster environment has on galaxy morphology. Therefore, we focus on clusters of mass $M_{\text{200c}}\sim10^{14-14.6}\MSUN$, the upper limit being set by the most massive cluster available in TNG100. There are 14 such clusters in TNG100 and 2 in TNG50, the latter with masses of $M_{\text{200c}}=1.8\times10^{14}\MSUN$ $M_{\text{200c}}=0.9\times10^{14}\MSUN$. Further details of the clusters are provided in Table \ref{tab:hostProps}. Stellar density maps of the two TNG50 clusters, the most massive ones in the simulation volume, are shown in Figs. \ref{fig:tng50_halo0} and \ref{fig:tng50_halo1} illustrating the immense diversity of galaxies in these systems. With stellar particle masses of $8.5\times10^{4}\MSUN$, these are currently the highest resolution simulations of such Virgo-mass clusters. The two clusters are resolved by over 133 and 54 million stellar particles, over 256 and 129 million gas cells, and over 415 and 205 million DM particles, respectively. Similar resolutions for such high mass hosts have only been previously achieved in state-of the art zoom-in cluster simulations such as Romulus-C \citep{Tremmel19} with gas particle masses of $\sim10^{5}\MSUN$ and the Hydrangea suite of simulations \citep{Bahe17} with baryon particle masses of $\sim10^{6}\MSUN$.

We first select galaxies that lie within the cluster's virial radius at $z=0$ and have stellar masses of $M_{*}>3\times10^{8}\MSUN$ in TNG50 and $M_{*}>5\times10^{9}\MSUN$ TNG100. Cluster centrals, as identified during the halo finding process, are excluded from this analysis. These criteria provide us with 226 cluster satellites out of 5898 luminous subhaloes and 90760 dark subhaloes in TNG50 and 528 cluster satellites out of 15084 luminous subhaloes and 93630 dark subhaloes in TNG100 (where luminous subhaloes are any subhaloes with non-zero stellar mass).

In a companion paper (Joshi et al. in prep.) we quantify the morphological mix of the $z=0$ cluster satellite populations of TNG100 and TNG50 and demonstrate that a morphology-density relation is an emergent feature of the IllustrisTNG model. The existence in the simulations of a dependence on cluster environment of the morphological diversity of galaxies and its similarity with observational constraints lend credibility to the quantitative results of the analysis that follows here. From Figs.~\ref{fig:tng50_halo0} and \ref{fig:tng50_halo1} such morphological diversity is apparent, with satellites appearing both as massive spheroidals -- in a few instances even equipped with their own shell-like signatures of recent interactions -- as well as razor-thin disc-like galaxies across masses. In fact, Figs.~\ref{fig:tng50_halo0} and \ref{fig:tng50_halo1} also provide indications for strong host-to-host variations, whereby the high-mass end of the satellite populations appears rather different in the two clusters: one is dominated by elliptical galaxies and shows overall indications of recent relaxation, the other features a number of massive disc satellites and overall a less spherical stellar mass distribution and a somewhat more dynamical state.

\subsection{Control parent sample} \label{sec:sampleControl}
To differentiate between the effect of environment and the secular evolution of our satellite sample, we require a sample of galaxies that evolves in parallel with the cluster satellite population and whose evolution is dominated by secular processes. We therefore define a control sample of galaxies mass-matched to the cluster satellites.

While it is possible to mass-match the galaxies at $z=0$, previous studies have shown that satellites and centrals exhibit significantly different evolution in stellar mass \citep[e.g. see][]{Mistani16}. Therefore, we select our control sample by mass-matching the cluster satellites at their time of accretion. `Accretion' is defined as the last snapshot before the galaxy \emph{first} becomes part of its \emph{final} host (i.e. the galaxy is identified as part of the $z=0$ host cluster's progenitor FOF group; the galaxy is well beyond the cluster virial radius at that time in most cases). 

We use the following procedure to define the control sample:
\begin{itemize}
    \item For each cluster satellite, we consider all galaxies that are centrals at $t_{\text{acc}}$, when the galaxy was accreted.
    \item Only galaxies that remain centrals up to $z=0$ are retained.
    \item Of these, we select the three galaxies that are the closest in stellar mass to the cluster satellite at $t_{\text{acc}}$.
\end{itemize}
Note that we do not remove duplicated galaxies from this sample. We thus have a control sample of 678 (548 unique) galaxies in TNG50 and 1584 (1216 unique) galaxies in TNG100. In later sections, we use the time since accretion of the cluster galaxies to study their evolution and compare them to the evolution of the control sample over the same timescales. Hence we define an `accretion' time for the control galaxies which is equal to the accretion time of the corresponding cluster satellites they were mass-matched to. \emph{We refer to this reference time for the control galaxies as an accretion time, for convenience.}

% ------------------------------------------------------------------------------------------------------------

\begin{figure*}
	\includegraphics[width=\linewidth]{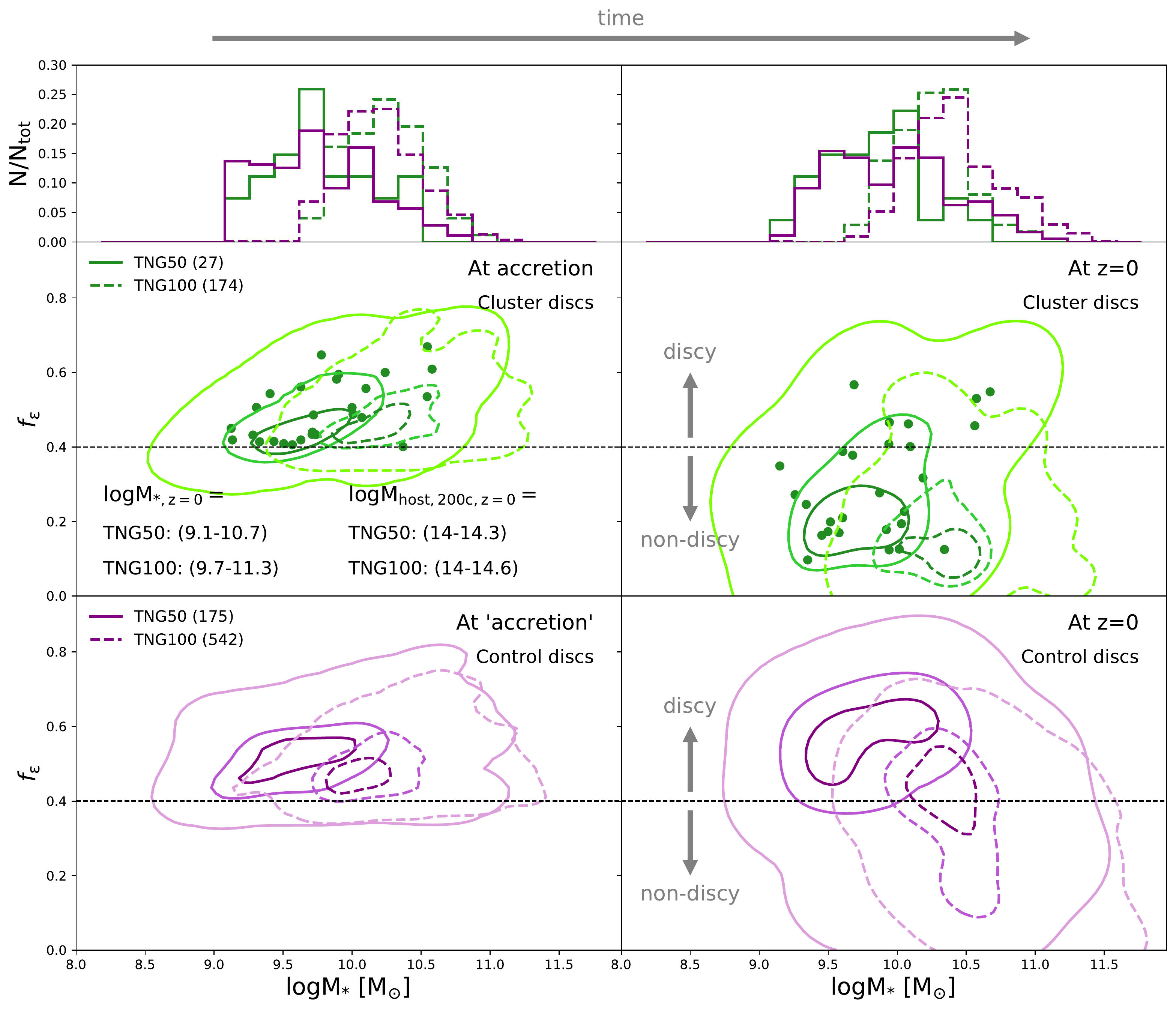}
	\caption{\textbf{Morphological changes across time due to environment} The contours show the distribution of circularity fraction i.e. disciness, and stellar mass for the discy cluster satellites (middle row) and discy control galaxies (bottom row), at the time of accretion (left panels) and at $z=0$ (right panels). Solid lines show the TNG50 sample, dashed lines the TNG100 sample. The contours enclose 20\%, 50\% and 90\% of the data within each sample. The top row provides the corresponding stellar mass distributions for all four subsamples. Note that the contours in the left panels only extend below $f_{\epsilon}<0.4$ due to the smoothing applied to generate the contours. For the TNG50 cluster disc sample, we also show individual points as green symbols. Nearly all cluster discs are non-discy by $z=0$, whereas a large portion of the control discs remains discy over the same timescales.} \label{fig:circVsMstar}
\end{figure*}

\begin{figure*}
    \includegraphics[width=\linewidth]{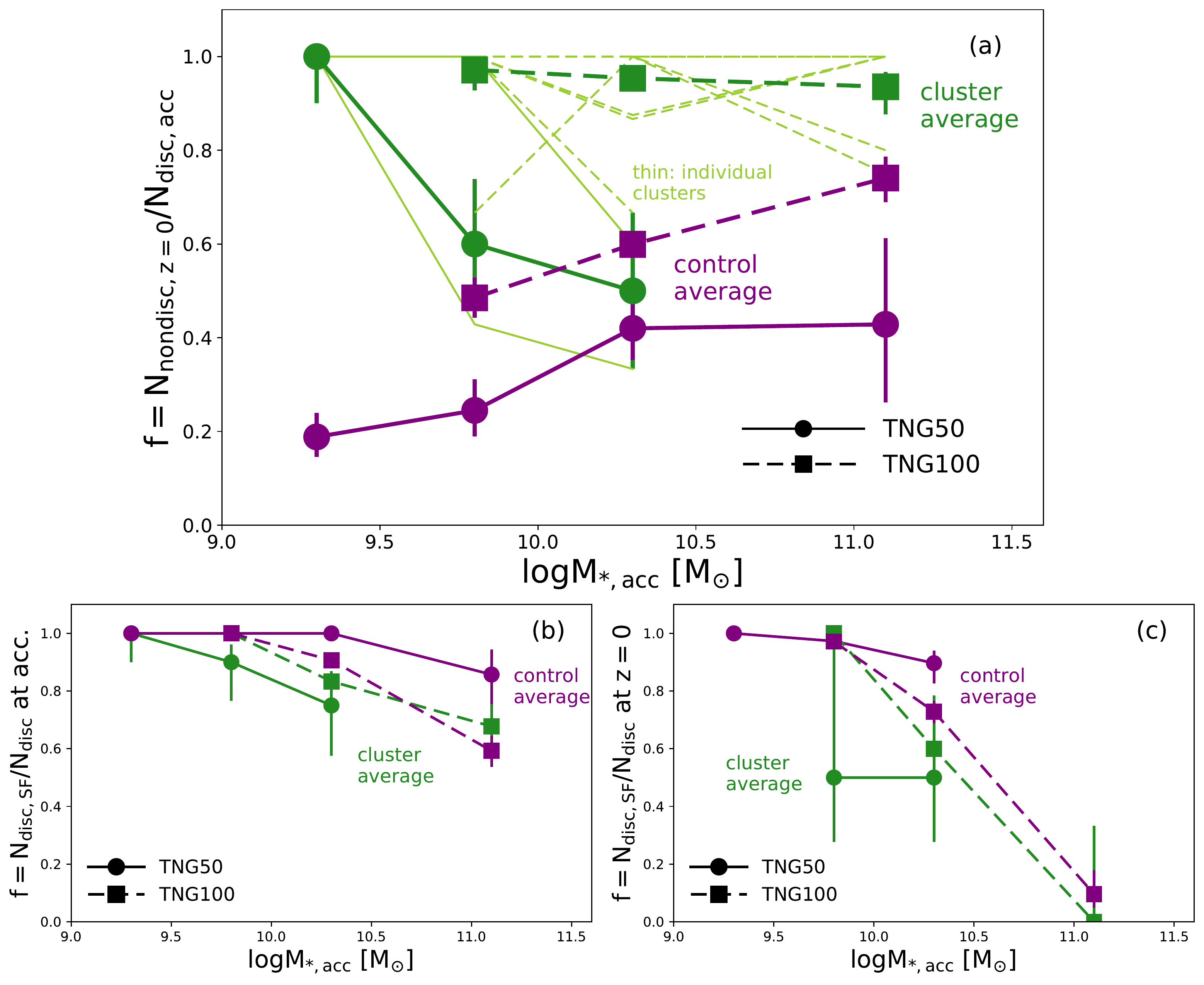}
    \caption{\emph{(a):} Fraction of cluster and control discs that have transformed to non-discs by $z=0$ in bins of stellar mass at accretion. Error bars show the Wilson score interval for each value. Note that for the control sample, only stellar mass bins with at least 5 galaxies are plotted. Thin, light green lines show the results for the cluster discs within each cluster separately. The fraction of cluster discs that do not survive to $z=0$ is always higher than control discs that do not survive. Moreover, the cluster and control discs show opposite trends with stellar mass, such that larger (smaller) fractions of more massive cluster (control) discs survive to $z=0$. \emph{(b) \& (c):} Fraction of cluster and control discs that are also star-forming, both at accretion (b) and at $z=0$ (c). Comparing the results at accretion and at $z=0$ at fixed stellar mass, there are relatively fewer star forming discs at $z=0$. The effect is largest for the most massive galaxies.} \label{fig:discDestructionFrac}
\end{figure*}

% ------------------------------------------------------------------------------------------------------------

\subsection{Cluster disc and control disc samples}
In this study, we only consider specific subsamples of galaxies from the parent cluster satellite and control populations defined in the previous subsections -- we select only those galaxies that were discs at the time of accretion (according to the stellar circularity fraction defined in Section~\ref{sec:methodsMorph}). As shown in \citet{TNG50Pillepich19}, disc morphologies are rare even amongst star-forming galaxies of masses $M_{*}\lesssim10^{9}\MSUN$, and this is not due to a resolution effect but rather is a physical finding. In fact, as shown in fig. 9 of \citet{TNG50Pillepich19}, the trends of disc fractions in TNG50 with stellar mass and redshift are in good agreement with observations down to stellar masses of $M_{\text{*}}\sim 10^{9}\MSUN$, where SDSS and CANDELS data are available. Hence, we focus on the mass range of $\log{M_{\text{*,acc}}}=9-11.6$ for the remainder of the paper (the upper limit being set by the most massive discy cluster satellite available in our sample). The mass ranges we quote in the rest of the paper are for the disc subsamples only. Therefore, we define a cluster disc and a control disc sample from each of the TNG50 and TNG100 parent samples, consisting of 27 cluster discs and 191 control discs (of which we exclude the 16 galaxies with $M_{\text{*,acc}}<10^{9}\MSUN$) in TNG50, and 174 cluster discs and 542 control discs in TNG100. In Tables \ref{tab:hostProps} and \ref{tab:sample}, we provide the sizes of each of these samples, divided by host cluster and in bins of galaxy stellar mass at the time of accretion respectively. Importantly, not all these galaxies will remain discs at $z=0$ (the focal point of this paper) and so in Tables \ref{tab:hostProps} and \ref{tab:sample} we also give the number of galaxies that remain discs through $z=0$. 

Table \ref{tab:hostProps} shows that in all clusters except CL17 in TNG100, fewer than half the cluster satellites are actually discs at accretion. Since we define accretion as the first time galaxies become part of the progenitor of their $z=0$ clusters, galaxies can have been preprocessed in smaller hosts before this accretion. In fact, about $52-66\%$ of all $z=0$ cluster satellites in both TNG50 and TNG100 have been preprocessed i.e. have been part of another group or cluster with a minimum mass of $M_{\text{200c}}>10^{12}\MSUN$ before accretion onto the progenitor of the cluster where we find them at $z=0$ (see also Donnari et al. in prep., for an extensive characterization of preprocessing in the IllustrisTNG simulations). Hence, a majority of the cluster satellites have already experienced some environmental effects before accretion onto their final hosts.

Because of group preprocessing, the morphological mix of future cluster satellites may already be different at accretion compared to the field and very few of the cluster satellites are actually discs even at accretion: $\sim21\%$ in TNG50 within the mass range of $\log{M_{\text{*,acc}}}=9-10.6$ and $\sim34\%$ in TNG100 within the mass range of $\log{M_{\text{*,acc}}}=9.6-11.6$. The corresponding initial disc fractions for the control galaxies in the same mass ranges are $\sim43\%$ and $\sim35\%$ respectively. These numbers are also determined by the specific assembly histories of the underlying hosts. For example, the most massive cluster in TNG50, CL0, has had two significant major mergers\footnote{Major mergers for cluster haloes are defined as a $>25\%$ increase in $M_{\text{200c}}$ between consecutive snapshots (corresponding to a virial mass ratio of $>1/4$ of the progenitors).} since $z=2$, the last one occurring at $z\sim0.36$. Hence, a large proportion of its satellites were preprocessed in hosts of considerable mass. On the other hand, CL1 of TNG50 has had no such major merger. Among the 14 hosts in TNG100, CL1, CL6 and CL9 have also had at least one merger since $z=1$ and an additional 1-3 major mergers between $z=1$ and $z=2$.

The distribution of accretion times of the cluster disc satellites is shown in Fig. \ref{fig:accTimesDist} in Appendix \ref{sec:clusterSampleStats}, along with the analogous timescales for the mass-matched control discs. Typically, the accretion time distributions of Virgo-like satellites that survive until $z=0$ are bimodal \citep[see also][]{Yun19, Engler20}, with $\sim85\%$ of the TNG50 cluster discs and $\sim65\%$ of the TNG100 cluster discs in our sample having fallen into their final $z=0$ cluster host at $z\leq1$.

% ------------------------------------------------------------------------------------------------------------

\begin{figure*}
	\includegraphics[width=\linewidth]{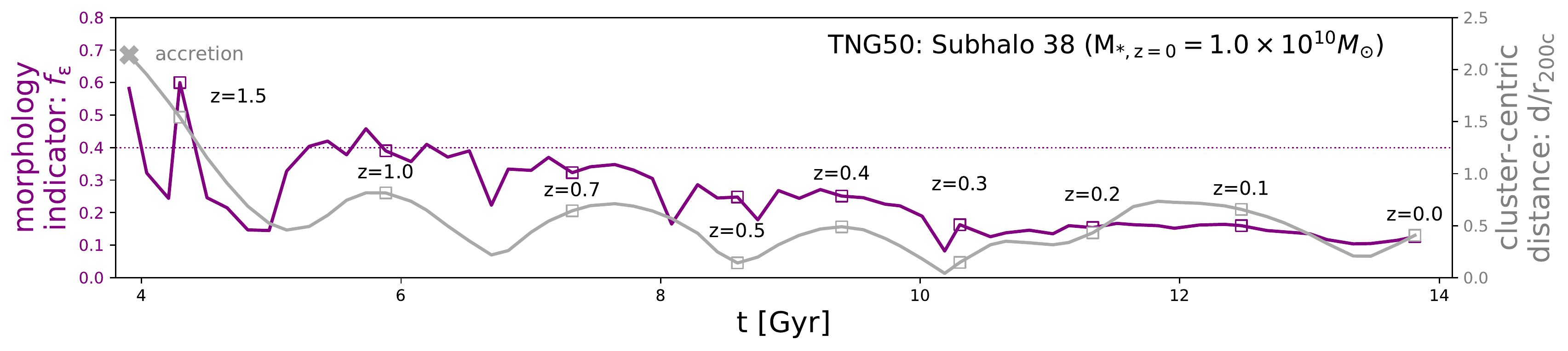}
	\includegraphics[width=\linewidth]{figures/gjoshi_clustergalaxies_histories_tng_2160_99_IDatz0-38_stars.pdf}
	
	\vspace{1cm}
	
	\includegraphics[width=\linewidth]{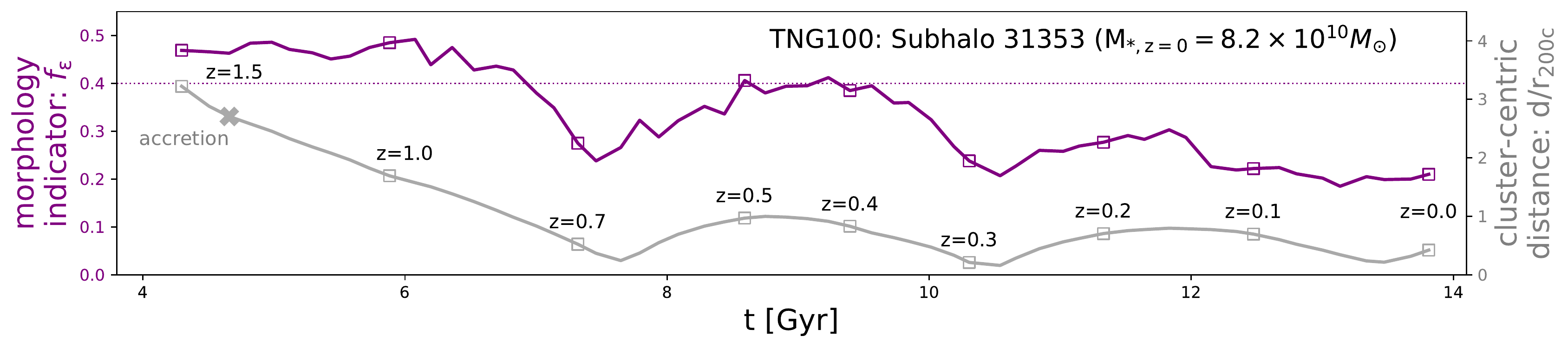}
	\includegraphics[width=\linewidth]{figures/gjoshi_clustergalaxies_histories_tng_1820_99_IDatz0-31353_stars.pdf}
	\caption{\textbf{Morphology evolution of two example cluster discs} from TNG50 (Rows 1-2) and TNG100 (Rows 3-4). \emph{Rows 1 \& 3:} Time evolution of the galaxy's morphology (purple), measured by the circularity fraction $\mathit{f_{\epsilon}}$, and its distance to its host cluster, normalized by the cluster's virial radius at the time (gray). \emph{Rows 2 \& 4:} Stamps showing the corresponding projected stellar light composite maps of the galaxies at the redshifts highlighted with square symbols in the panels above. Images are for the JWST NIRCam f200W, f115W, and F070W filters (rest-frame); we neglect any dust effects and include all stellar light within the given projection from gravitationally-bound stars. Face-on projections are shown on top, edge-on projections on the bottom. Every panel is 40 physical kpc on a side.} \label{fig:clusterDiscEvolMaps}
\end{figure*}

\section{Evolution of cluster discs} \label{sec:clusterDiscs}

We first examine whether the morphologies of the cluster samples differ from those of the control samples and, therefore, whether the cluster environment actually affects their morphological transformation.

In Fig. \ref{fig:circVsMstar}, we first look at the distributions of circularity fraction $\mathit{f}_{\epsilon}$ and stellar mass $M_{*}$ for each of the disc samples. The left panels show the initial distributions at the time of accretion. All four distributions are broadly similar apart from the TNG50 samples extending to lower masses by selection. The results are starkly different at $z=0$, shown in the right panels. While most of the cluster discs undergo a strong reduction in their stellar circularity fraction, to the point that most of them become non-discy in the cluster environment, a majority of the control discs remain discy over the same timescales, i.e. with high circularity fractions. These changes are also accompanied by growth in stellar mass for the control discs, while the cluster discs show no increase in stellar mass. We explore the relation between mass gain/loss and change in morphology in the next section. 

To our knowledge, this is the first time that the morphological transformation of cluster galaxies has been demonstrated to occur with fully cosmological simulations of the formation of galaxies; within this framework we are able to show that the cluster environment effectively imparts differential effects with respect to morphological transformations that affect galaxies that are centrals or in lower-density environments. To further quantify the population-wide changes seen in Fig. \ref{fig:circVsMstar}, Fig. \ref{fig:discDestructionFrac}(a) shows the fraction of cluster (green) and control (magenta) galaxies that, being discy at accretion ($\mathit{f}_{\epsilon}\text(acc.)>0.4$), become non-discy at $z=0$ ($\mathit{f}_{\epsilon}\text(z=0)\leq0.4$), as a function of stellar mass at accretion. Thick curves and large symbols denote averages across the TNG50 (solid) and TNG100 (dashed) hosts and satellite populations; thin curves indicate fractions on a cluster-by-cluster basis, with 9 of the 14 TNG100 clusters overlapping on the $f=1$ line and with several of the lower-mass clusters not having discs in the most massive bin. The bins of stellar mass at accretion were chosen to contain approximately equal numbers of TNG50 cluster discs, which is the smallest of our four samples, in the lower mass bins. The highest mass bin includes all high mass cluster discs in TNG100.

Fig. \ref{fig:discDestructionFrac} shows two key trends. Firstly, the fraction of cluster discs that become non-discy is always higher than that for the corresponding control sample, as seen in the Fig. \ref{fig:circVsMstar}. Secondly, the fraction shows a rising trend with initial galaxy stellar mass for the control samples in both TNG50 and TNG100, while the TNG50 cluster sample shows a decreasing trend with increasing galaxy mass. The trend for the TNG100 cluster discs is milder, nearly consistent with having no trend on average. This generally indicates that the higher-mass cluster discs seem to be more resilient to any environmental influence they experience compared to lower-mass satellites, while for the control discs, their predominately secular evolution drives higher-mass discs to become less discy. In other words, in the field, high mass galaxies ($\gtrsim$ a few $10^{10}\MSUN$) have a higher chance of becoming less discy or non-discs with time compared to lower mass galaxies. It is important to note that the fractions shown in Fig. \ref{fig:discDestructionFrac} depict the proportion of galaxies that were discs at accretion and are non-discs at $z=0$, not the total non-disc fractions around clusters. In principle, this would seem to indicate that, especially in the case of TNG100, the cluster disc fractions are significantly lower than observed. This is partly due to differences in morphology measurements between simulations and observations and partly due to selection effects. In Joshi et al. in prep., we will explore these issues in more detail and show that the disc fractions in various environments in IllustrisTNG at $z=0$ are in fairly good agreement with observations.

Differences between TNG100 and TNG50 are evident in the top panel of Fig. \ref{fig:discDestructionFrac}, with the fraction of non-discs at $z=0$ always being lower for TNG50 than TNG100. We believe that this disparity is driven by a combination of different numerical resolutions and different host realizations. Firstly, in the control samples (magenta curves), the resilience to morphological change is higher in the higher-resolution TNG50 than in TNG100, by about 20 percentage points at all masses; this is consistent with a more general dependence of stellar morphology on numerical resolution. For example, \citet{TNG50Pillepich19} have shown that the resolution of TNG50 is required to capture thin stellar discs, especially at lower masses, and that disc galaxies at TNG100 resolution are generally puffier than those in TNG50 at most times. The effects of resolution are evident on the cluster satellites as well; however, they are not the only culprit for the diversity between the TNG100 and TNG50 cluster samples. Thin, light green curves in Fig.~\ref{fig:discDestructionFrac}(a) show the results for each cluster separately, solid for TNG50 and dashed for TNG100. There is a significant host-to-host variation in the trends, even among the TNG100 hosts. One of the TNG100 cluster samples (in CL15) shows a similar resilience to become non-discy as one of the TNG50 clusters (CL0). We have checked that the different trends between the TNG50 and the TNG100 clusters seem not to be driven by host mass (the TNG50 hosts occupying the low-mass end of the TNG100 host mass distribution). Instead, a key factor in driving these differences is the distribution of accretion times of the disc galaxies in the different hosts, in turn driven by somewhat diverse cluster mass assembly histories. As can be seen from Fig. \ref{fig:accTimesDist}, the distributions of the TNG50 disc samples are significantly different from those of the TNG100 samples -- TNG100 cluster discs of all masses were accreted roughly at $z\sim1$ on average, $\sim2-4$ Gyr earlier than the TNG50 cluster discs which were accreted at $z\sim0.5$ or even later. In particular, we have checked the infall times of all individual galaxies and in the second most massive cluster of TNG50 (CL1), the host exhibiting the lowest fraction of disc galaxies that become non-discs, the average infall times of its more massive galaxies ($\gtrsim 10^{10}\MSUN$) is $\sim11.2$ Gyr ($z\sim0.21$), in comparison to an average of $\sim7.0$ Gyr ($z\sim0.96$) across all other TNG50 and TNG100 clusters -- we discuss the connections between morphological transformation and accretion times in the next Section. Overall, with only two clusters in the TNG50 sample, each with a unique assembly history, we do not have a fully representative probe of morphological transformation in cluster environments in TNG50. Therefore low number statistics of the clusters, along with numerical resolution effects, are responsible for the differences between TNG50 and TNG100.

In the lower panels of Fig. \ref{fig:discDestructionFrac}, we also compare the correspondence between morphology and star-formation status, by examining the fraction of cluster and control discs, at the time of accretion (panel (b)) and at $z=0$ (panel (c)), that are also star-forming. We employ a simple definition of quenching here, designating galaxies with sSFR$>10^{-11}\,\text{yr}^{-1}$ as quenched. At the time of `accretion', a majority of disc galaxies are also star forming. However, in TNG100 the fraction of quiescent (red) discs at accretion increases with galaxy stellar mass, both in the clusters and in the field, whereas nearly all the TNG50 control discs are star-forming. At $z=0$, these trends in TNG100 are even stronger, i.e. the fraction of TNG100 galaxies that are both disc and star forming decreases strongly with galaxy mass, whereas this does not appear to be the case for TNG50 (with the exception of the TNG50 cluster discs, where the number of galaxies is too small to discern any significant correlation between star-forming fraction and stellar mass). Note that in the right panel of Fig. \ref{fig:discDestructionFrac}, we only consider galaxies that are discs both at accretion and at $z=0$. Hence, a less biased and more in-depth study is required to pin down the effects of numerical resolution on the occurrence of red disc galaxies within the IllustrisTNG model \citep[see][for a discussion on red discs at lower redshifts in TNG100]{Tacchella19,RodriguezGomez19}.

The change in morphology for two example cluster discs is shown in Fig. \ref{fig:clusterDiscEvolMaps}. We select one galaxy from the TNG50 and TNG100 cluster disc samples, respectively, and show the time evolution of its morphology and its distance from the its host cluster centre, normalized by the host's virial radius at the time. For each galaxy, we also show stamps of the projected stellar light composites, face-on projections on top and edge-on projections in the bottom, at select redshifts; these are highlighted with the square symbols along the evolutionary tracks. While the overall evolution in visual morphology is consistent with the evolution of the circularity fraction, the two properties are not always perfectly correlated -- low values of $\mathit{f}_{\epsilon}$ do not always translate to visually elliptical morphologies. We have confirmed that there is a diversity in the final visual morphologies of the cluster discs. Especially for the more massive galaxies, morphological transformation can come in the form of the dissipation of spiral arm structure, the strengthening of a bar structure or the appearance of a boxy distribution in the face-on projections. At the low mass end, the morphological transformations generally result in mild disc thickening as well as more centrally concentrated stellar distributions. We examine the drivers of some of this diversity of morphological transformation in later sections and we showcase more examples for the time evolution of TNG galaxies in Figures~\ref{fig:egStampsTNG50} and \ref{fig:egStampsTNG100}.

\begin{figure*}
	\includegraphics[width=\linewidth]{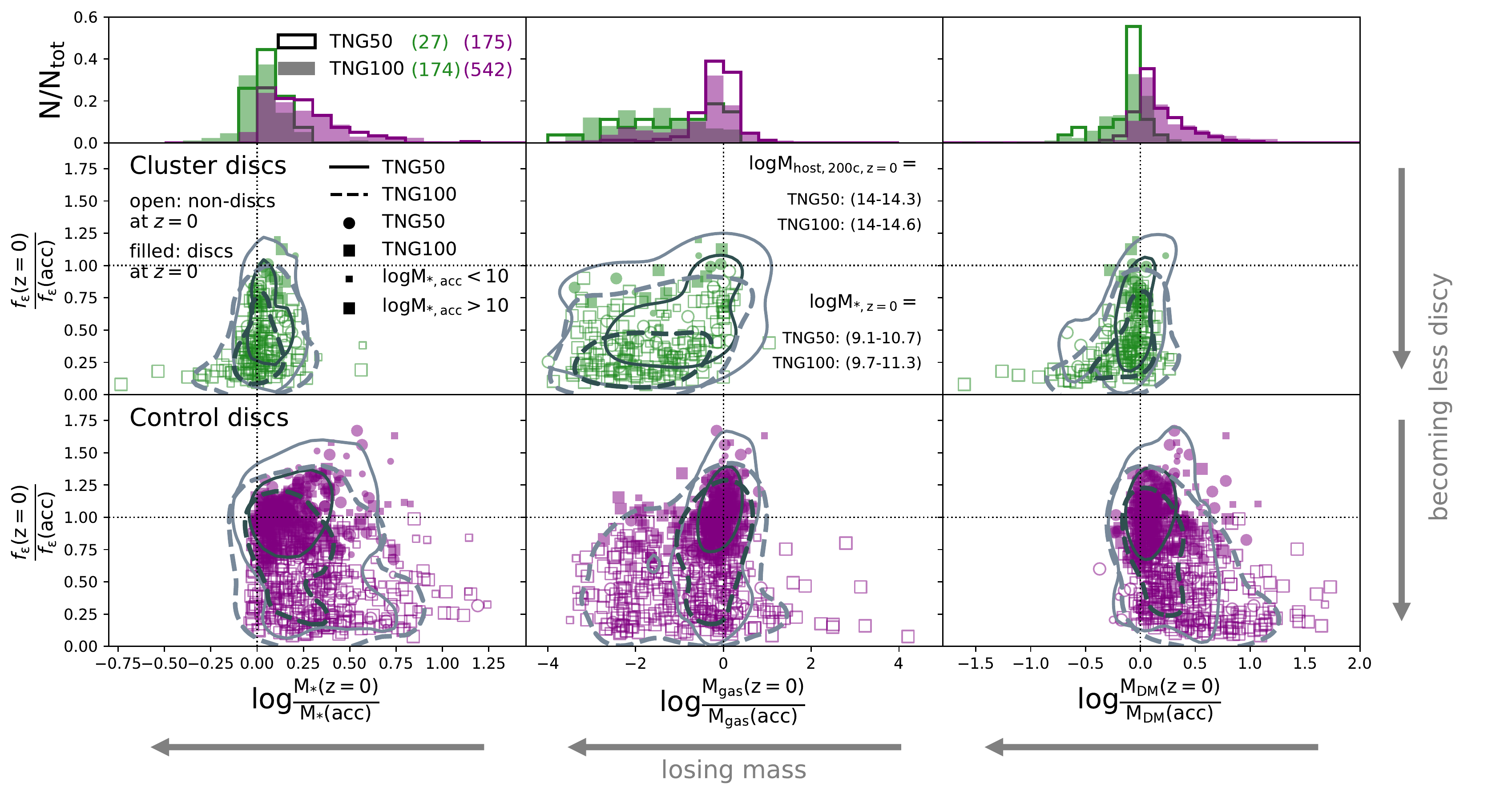}
	\caption{\textbf{Change in morphology versus changes in stellar mass (left column), gas mass (middle column) and dark matter mass (right column) since accretion to present day.} Top row panels show the distributions of the corresponding galaxy masses, normalized by total sample size. Note that in the middle column of panels, a significant fraction of cluster discs and a small number of control discs have zero gas masses at $z=0$ and a small number even at accretion. For such galaxies, we use the last non-zero gas mass they had before $z=0$ or accretion, as needed. Filled and open symbols show galaxies that are discs and non-discs at $z=0$ respectively. The size of the symbol denotes the $M_{\text{*,acc}}$ of the galaxy. Gray contours enclosing 50\% and 90\% of the data within each subsample are added to show the overall distribution of galaxies in these planes. The transformation to non-disc morphology is accompanied by significant loss of gas mass for all samples. However, for cluster discs, the transformation occurs along with loss of DM mass and no little gain in or mild loss of stellar mass, whereas for the control discs, it occurs along with significant gain in stellar and DM mass.} \label{fig:changeCircVsChangeMass}
\end{figure*}

\begin{figure}
	\includegraphics[width=\linewidth]{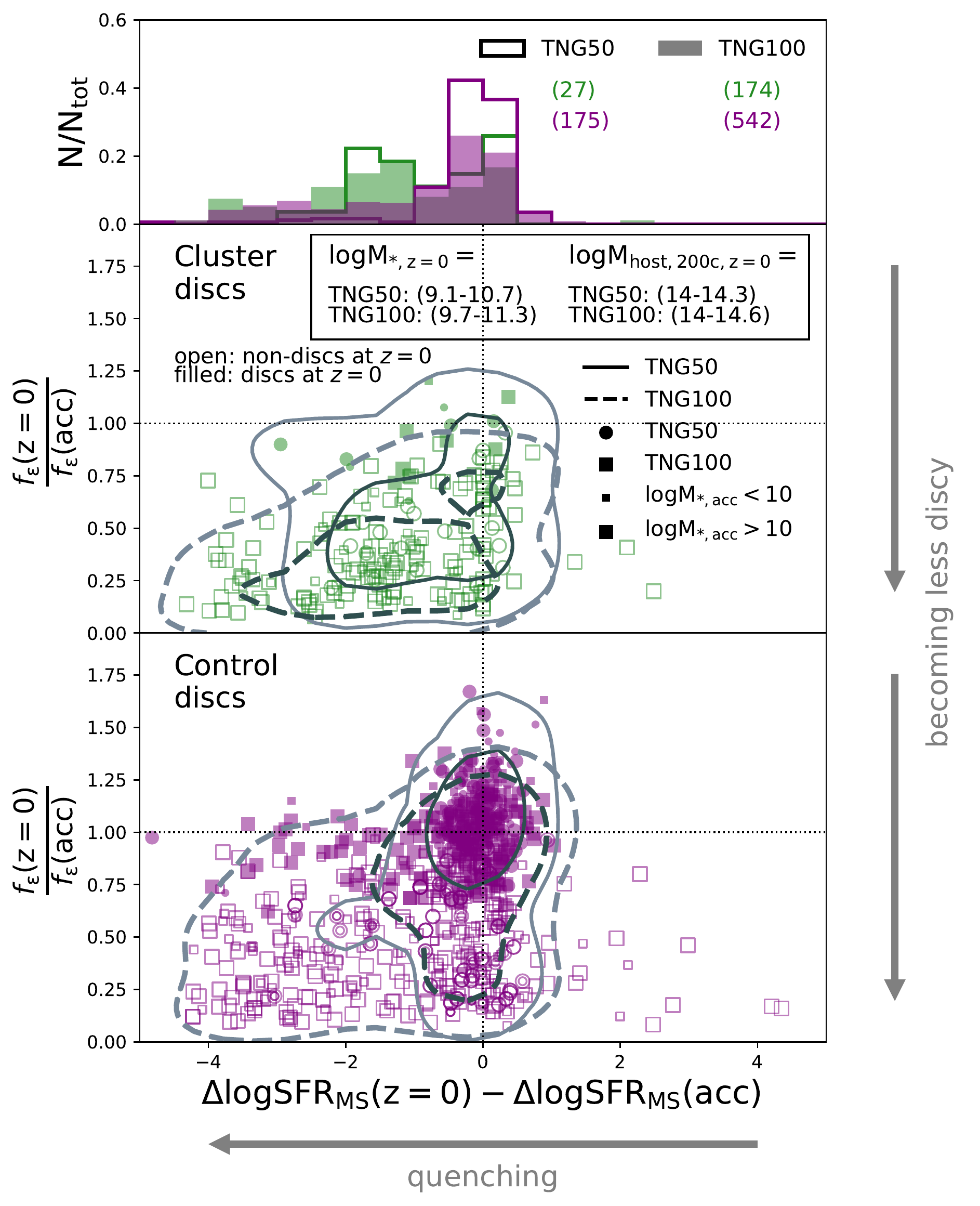}
	\caption{\textbf{Morphological change vs. change in SFR.} Change in morphology versus change in  $\Delta\log{\text{SFR}}_{MS}$, i.e. the distance from the SF main sequence, since accretion to present day. $\Delta\log{\text{SFR}}_{MS} = \log{\text{SFR}(z)}-(MS(z))$ is the difference in $\log{\text{SFR}}$ of the galaxy and the fit to the SF main sequence $MS$ at the given redshift; see text for further details. Cluster discs are shown in the middle panel, control discs in the bottom panel. As in Fig. \ref{fig:changeCircVsChangeMass} for gas mass, for galaxies that have zero SFRs at $z=0$ or at accretion, the values shown are their last non-zero SFRs. Filled and open symbols show galaxies that are discs and non-discs at $z=0$ respectively. The size of the symbol denotes the $M_{\text{*,acc}}$ of the galaxy. Gray contours enclosing 50\% and 90\% of the data within each subsample are added to show the overall distribution of galaxies in these planes. The transformation to non-disc morphology is accompanied by a significant decrease in SFR for all samples.} \label{fig:changeCircVsChangeDelSsfr}
\end{figure}

\begin{figure}
	\includegraphics[width=\linewidth]{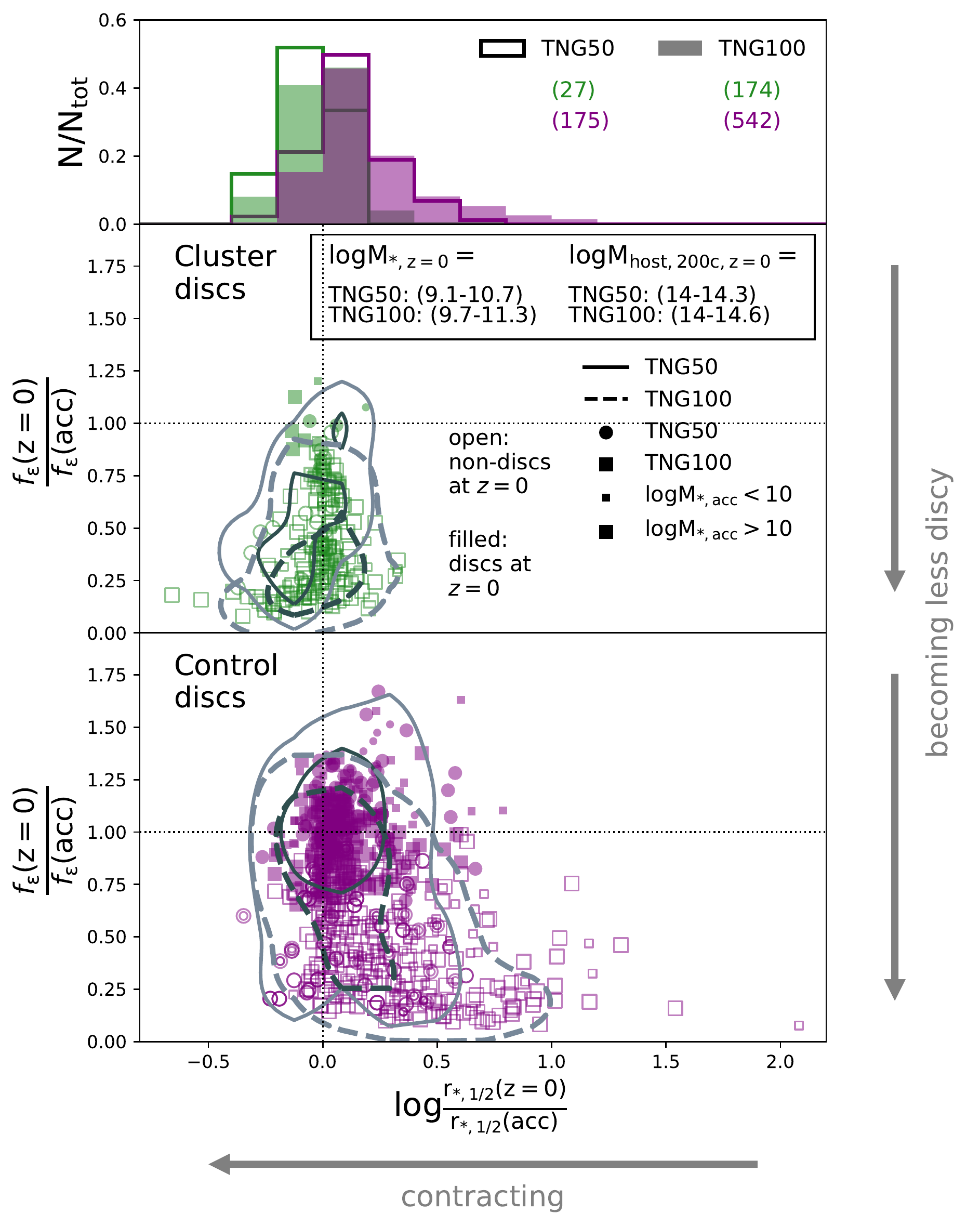}
	\caption{\textbf{Morphological change vs. change in stellar size.} Change in morphology versus change in the (physical) stellar half mass radius $r_{\text{*,1/2}}$ since accretion to present day. Cluster discs are shown in the middle panel, control discs in the bottom panel. Filled and open symbols show galaxies that are discs and non-discs at $z=0$ respectively. The size of the symbol denotes the $M_{\text{*,acc}}$ of the galaxy. Gray contours enclosing 50\% and 90\% of the data within each subsample are added to show the overall distribution of galaxies in these planes. Cluster discs become non-discs along with a moderate increase or decrease in stellar size, whereas control discs do so while showing significant growth in stellar size.} \label{fig:changeCircVsChangeRhalf}
\end{figure}

\subsection{Comparison to degree of change in other properties}

We further examine the degree of change in morphology for individual galaxies compared to the change in other galaxy properties. We focus on the mass of various components, the star-formation rate, and the galaxy stellar sizes, from accretion to present day. 

\subsubsection{Mass content}
In Fig. \ref{fig:changeCircVsChangeMass}, we show the ratio of the circularity fraction at accretion to that at $z=0$, versus the corresponding ratio of stellar mass (left column), gas mass (middle column) and DM mass (right column). The cluster and control discs are shown in the middle and bottom row respectively, while the top row provides the relative distributions of the corresponding galaxy mass ratios, normalized to the total number of objects in each simulation. The data points represent individual galaxies; larger symbols indicate higher-mass galaxies ($\geq10^{10}\MSUN$). A large fraction of the cluster discs ($\sim33\%$ for TNG50, $\sim87\%$ for TNG100) and a few TNG100 control discs have zero gas masses at $z=0$; a few galaxies even have zero gas masses at accretion. For such galaxies, we use the last non-zero gas mass value they had before $z=0$ or before accretion, as needed. 

The lower row of Fig. \ref{fig:changeCircVsChangeMass} shows that those control discs that become less discy by $z=0$ tend to also grow substantially in stellar and DM mass, while losing large portions if not all of their gas mass. However, this is a small fraction of the control disc samples as it is dominated by $\gtrsim 10^{10}\MSUN$ galaxies (larger symbols); the majority of control discs (lower-mass galaxies) remain discy to $z=0$ and these galaxies have seen moderate growth in their stellar and DM mass and less extreme loss, or indeed growth, of gas mass compared to the ones that become less discy. The cluster discs experience much lower growth in stellar mass and DM mass, in fact a significant portion lose mass in DM (and some also lose stellar mass); nearly all of them lose most if not all their gas mass. Cluster discs that have lost significant amounts of stellar and DM mass since accretion are also those that have become significantly less discy.

Thus, the change to less discy morphologies is accompanied by significant loss of gas mass for both cluster and control discs. However, while cluster discs become non-discy accompanied by a \emph{loss} of mass (or reduction in mass growth) in the stellar and DM components, control discs do so along with a \emph{growth} in stellar and DM mass.

\subsubsection{Star formation rate}
Fig.~\ref{fig:changeCircVsChangeDelSsfr} similarly shows the change in circularity fraction versus the change in SFR of the galaxies. As noted in the previous section, the great majority of the cluster and control disc galaxies are also star-forming at accretion. To determine if galaxies are being quenched as they become less discy, we measure their logarithmic distance from the locus of the star-forming main sequence (MS) of all galaxies at the given redshift. Following the procedure of \citet{Donnari19}, the main sequence is determined by first measuring the median SFR in 0.2 dex bins of $\log{M_{*}}$ within the range 9-10.2 and then calculating a linear fit in the $\log{\text{SFR}}$-$\log{M_{*}}$ plane. The process is iterated by successively removing galaxies beyond $2\sigma$ from the current main sequence fit, until the fit parameters converge to within 1\%.  As with gas mass in Fig. \ref{fig:changeCircVsChangeMass}, for galaxies with zero SFRs at $z=0$ or at accretion, we instead plot their last non-zero SFRs. Note that the number of galaxies that have zero SFRs are higher than those that have zero gas masses, in all samples and both at $z=0$ and at accretion. 

For both cluster and control discs, galaxies that become less discy by $z=0$ are likely to be quenched or have significantly reduced SFRs. We further discuss the connection between quenching and morphological transformation in Section \ref{sec:discussionSFRvsFcirc}. The $z=0$ distributions of the two disc populations (top panel of Fig.~\ref{fig:changeCircVsChangeDelSsfr}) are consistent with the majority of the cluster discs becoming quenched and most of the control discs remaining star-forming, although with the latter exhibiting a strong tail towards quenching for those galaxies that become non-discy and dominated by higher-mass galaxies ($\gtrsim 10^{10}\MSUN$, larger symbols).

\subsubsection{Stellar size}
Finally, we also compare the change in circularity fraction with the change in the galaxies' sizes, namely their stellar half mass radii $r_{\text{*,1/2}}$, in Fig. \ref{fig:changeCircVsChangeRhalf}. We have confirmed that there is no significant difference in the initial distributions (i.e. at accretion) of stellar sizes themselves, between the cluster and control samples. For the cluster discs, the change to non-disc morphology may be accompanied by a moderate decrease or increase in the physical stellar size of the galaxy, by up to a factor of $\sim3$ at most. The control discs show a stark contrast to this trend, with the majority of control discs growing in stellar size with time by a factor of many, if not more than one order of magnitude, especially those that become non-discy. The latter are in fact dominated by massive galaxies, which would be expected since largely, massive control galaxies necessarily reside at the centres of massive DM haloes where they can accrete vast amounts of stars that are stripped from incoming and merging satellite galaxies \citep[e.g.][]{RodriguezGomez16}. The stellar size growth we find here is qualitatively consistent with that observed for massive quiescent galaxies \citep{VanDerWel14} and explained with cosmological simulations by the accumulation of ex-situ i.e. accreted stars from minor mergers in the outskirts of galaxies \citep[][]{Oser12, RodriguezGomez16}. It should be noted that the growth (contraction) of the stellar half mass radius can be a natural consequence of the growth (loss) of stellar mass, in which case the results of Figs. \ref{fig:changeCircVsChangeRhalf} and \ref{fig:changeCircVsChangeMass} would be degenerate. However, we have confirmed that these results hold even when considering the stellar surface density within twice the stellar half mass radius $\Sigma_{*}=M_{*}/\left(4\pi (2\,r_{*,1/2})^{2}\right)$. Control discs show larger changes to non-disc morphologies while becoming significantly more diffuse, while cluster discs may do so while becoming moderately more diffuse or more concentrated.

% ------------------------------------------------------------------------------------------------------------

\begin{figure*}
    \includegraphics[width=\linewidth]{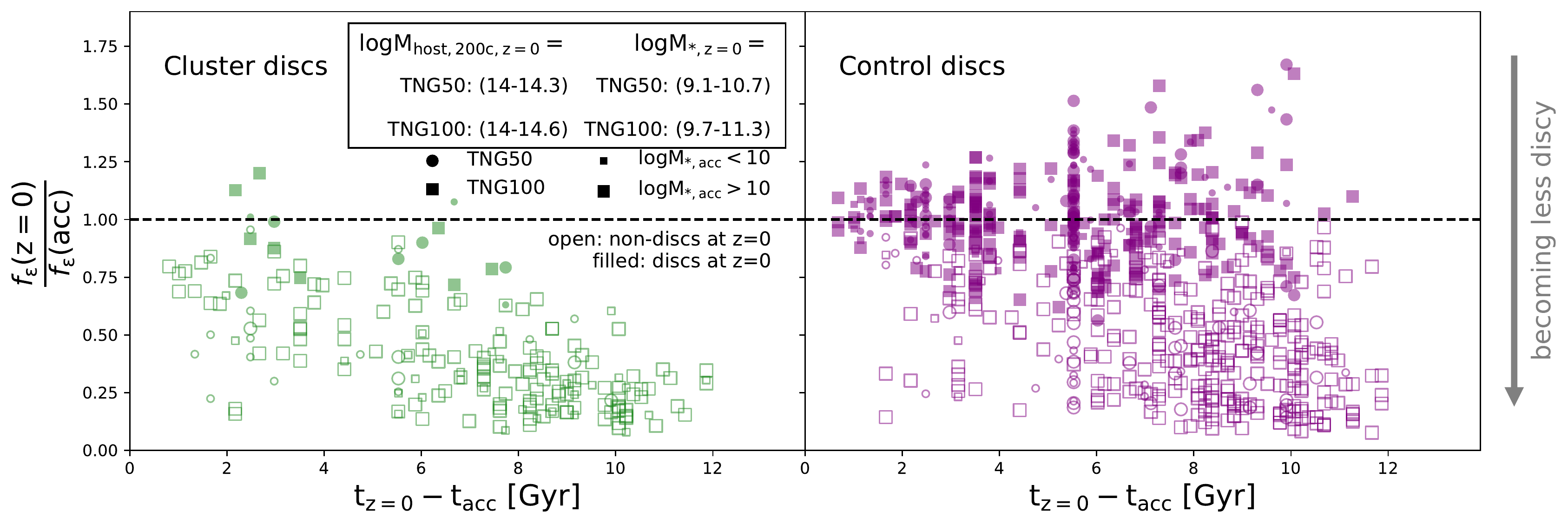}
    \includegraphics[width=0.915\linewidth]{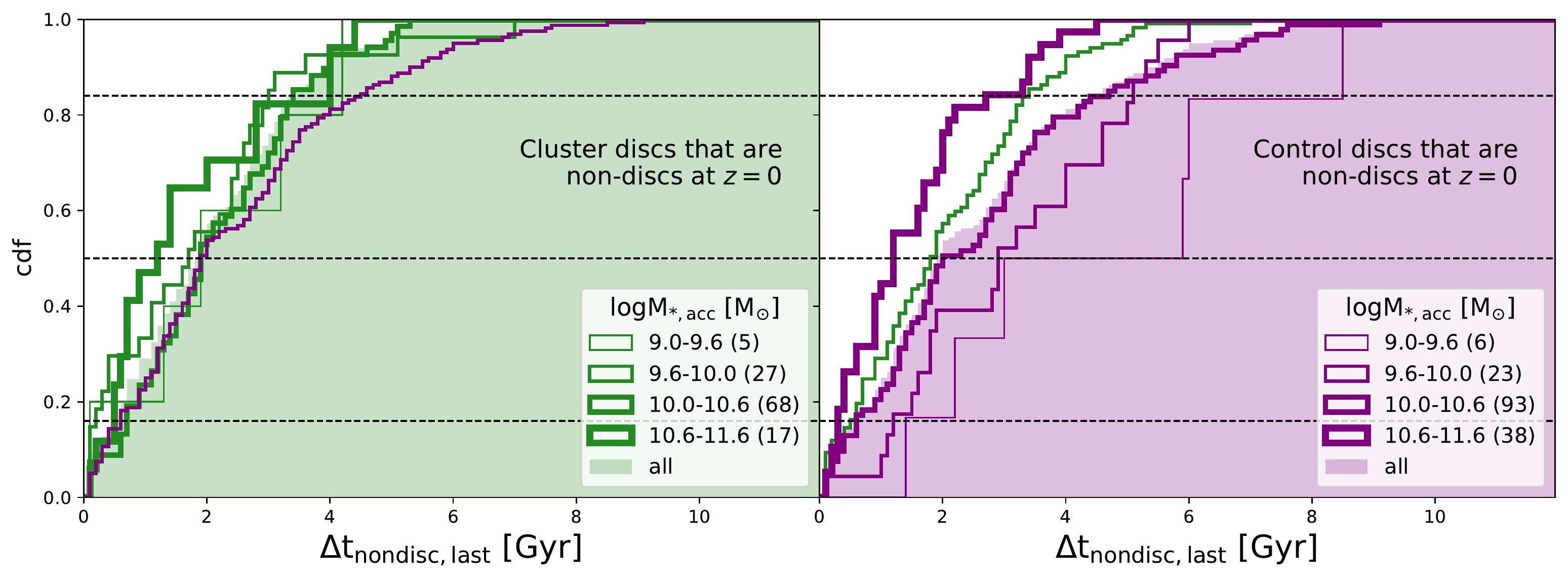}
	\caption{\textbf{Timescales of morphological change.} \emph{Top:} Change in morphology versus time since accretion to present day for the cluster discs (left) and control discs (right), where $t$ is cosmic time. The `accretion' time for the control galaxy is the accretion time of the corresponding cluster galaxy for which it was selected. Galaxies that remain discs at $z=0$ and that are non-discs at $z=0$ are shown by filled symbols and open symbols respectively, to emphasize the differences in the two populations. The size of the symbol denotes the $M_{\text{*,acc}}$ of the galaxy. The longer a cluster disc is in a cluster environment, the greater is its degree of transformation to a non-discy morphology. Although the degree of transformation of the control discs is also greater with time, there is considerable scatter in the relation with a significant fraction of control discs remaining discy over timescales of 8-10 Gyr. \emph{Bottom:} Distribution of time to become non-discs for the cluster (left) and control (right) disc samples. $\Delta t_{\text{nondisc,last}}$ is the difference between the accretion time and the last occurrence of a galaxy's morphology transitioning from discy ($\mathit{f}_{\epsilon}>0.4$) to non-discy ($\mathit{f}_{\epsilon}\leq0.4$). Shaded regions show the cumulative distribution function (CDF) for the entire disc sample, combining the TNG50 and TNG100 data, while individual lines show the distributions in bins of stellar mass at accretion. In the right (left) panel, we recreate the results for all galaxies from the left (right) panel in green (purple), to ease comparison between the cluster and control samples. While the time interval of morphological transformation of the cluster disc sample shows little dependence on initial stellar mass, massive control discs ($\log{M_{\text{*,acc}}}=10.6-11.6$) are seen to evolve significantly more rapidly than lower mass control discs. The median timescales of transformation are similar between the cluster and control samples, but the control galaxies exhibit a broader range of $\Delta t_{\text{nondisc,last}}$, especially at lower masses.} \label{fig:times}
\end{figure*}

\section{Timescale of morphological transformations} \label{sec:rates}

One key factor that can determine the degree to which a galaxy is affected by a group/cluster environment is the time it has spent there since accretion. Whether such a correlation exists can be explored through two measurements -- (i) the total degree of morphological change as a function of total time spent in the cluster and (ii) the time required for the discs to become non-discy after accretion. We examine both of these in Fig. \ref{fig:times}.

The top panels of Fig. \ref{fig:times} show the ratios between the circularity fractions at $z=0$ and at accretion, versus the time since accretion to present day. It is evident that the degree of change in circularity fraction for the cluster discs (left) is larger the longer the time spent within the cluster, with most of them having spent over 6 Gyr in their host clusters (i.e. accreted at $z\gtrsim0.6$) and having become significantly less discy. Over the same time periods, the control discs (right) also show a broadly similar correlation between change in morphology and timescale, but a significant portion of the galaxies remains discy even over 8-10 Gyr timescales (i.e. since $z\sim1-1.7$). Among the galaxies that remain discy at $z=0$ (filled symbols), we count 11 cluster discs (4 in TNG50, 7 in TNG100) with stellar mass $M_{\text{*,acc}}>10^{10}\MSUN$. Two of them are satellites of the second most massive host in TNG50 (CL1) and have fallen into the cluster relatively recently, on average $\sim2.6$ Gyr ago ($z\sim0.21$); in fact all 11 were accreted less than $\sim4.4$ Gyr ago ($z\sim0.4$). This compares to an average accretion time of $\sim6.7$ Gyr ago ($z\sim0.74$) for all massive disc galaxies across the clusters of TNG50 and TNG100, $\sim42\%$ being accreted at $z>1$ and $\sim4\%$ accreted at $z>2$. Thus, it appears that the reason that the fraction of massive cluster discs that transform to non-discs is lower than typical for CL1 in Fig.~\ref{fig:discDestructionFrac} (lowest thin green curve) is that they have spent shorter times within the cluster compared to the other cluster discs.

To understand the rate at which these changes occur, we explore the second of the previously mentioned measurements, the time to become non-discy, $\Delta t_{\text{nondisc,last}}$ (see Section \ref{sec:methodsTimescales} for definition), in the bottom panels of Fig. \ref{fig:times}. Only galaxies that are non-discs at $z=0$ are considered for this analysis, giving us a sample of 19 cluster discs and 49 control discs in TNG50 and 166 cluster discs and 320 control discs in TNG100. The results are shown in the bottom of Fig. \ref{fig:times}, as cumulative distribution functions (CDF), combining the TNG50 and TNG100 samples. The shaded regions show the full cluster (left) and control (right) disc samples (these are recreated in each panel for ease of comparison). Additionally, the various lines show the CDFs for bins of stellar mass at accretion. The left panel shows that the cluster discs become non-discy at similar rates irrespective of their stellar mass at accretion. The median time to become non-discy, $\Delta t_{\text{nondisc,last}}$, is about $2.0^{+2.1}_{-1.3}$ Gyr (ranges are 16$^{\rm th}$-84$^{\rm th}$ percentiles) after accretion for all but the lowest mass bin, where transformations are slightly faster with the median time being $1.5^{+1.4}_{-1.0}$ Gyr -- note that this lowest mass bin only contains TNG50 discs and hence has only 9 galaxies. The distribution of the times to become non-discs after infall are however quite broad; the corresponding $90^{\rm th}$ percentile values are approximately $4.6$ Gyr and $3.5$ Gyr, respectively.

In contrast, the right panel of Fig. \ref{fig:times} shows that the rate of morphological transformation of the control discs does exhibit a marked dependence on initial stellar mass, with the most massive galaxies evolving more rapidly compared to lower mass ones. The median $\Delta t_{\text{nondisc,last}}$ for the $\log{M_{\text{*,acc}}}=9-9.6$, $9.6-10$, $10-10.6$, and $10.6-11.6$ mass bins are $4.3^{+2.2}_{-2.3}$, $4.1^{+2.7}_{-2.4}$, $3.2^{+3.2}_{-2.2}$, and $1.6^{+2.1}_{-1.1}$ Gyr,  respectively; the corresponding 90th percentiles values of $\Delta t_{\text{nondisc,last}}$ are approximately $7.7$, $7.4$, $7.0$ and $4.0$ Gyr. Therefore, the cluster environment acts to remove the dependence on stellar mass of the timescales to become non-discs that would normally exist in the field, since all cluster discs become non-discy on timescales that are similar to those of the massive ($\gtrsim 10^{10.6}\MSUN$) control galaxies, about 2 Gyr on average (0.5-4 Gyr within one sigma). We have confirmed that if we had measured the \emph{first} time the galaxies become non-discs instead of the \emph{last}, the timescales for all subsamples are shorter as expected. The median $\Delta t_{\text{nondisc,first}}\sim1$ Gyr for both cluster and control discs, regardless of stellar mass, but with the control samples having a much broader range of timescales. However, the change in morphology can be stochastic over short timescales, and therefore, this measure is prone to significant noise; the last time the discs transform to non-discs, albeit also somewhat susceptible to this stochasticity, is better able to characterize the transformation timescale. Had we chosen to characterize the time to become non-discs as the time when the galaxy's circularity fraction last crosses the $\mathit{f}_{\epsilon}=0.2$ threshold rather than 0.4, these estimates would be longer by $\sim2.2$ Gyr for the cluster discs and $\sim1.5$ Gyr for the control discs. Note that when using this stricter definition for the timescale to become non-discy, the cluster and control discs show similar trends with mass such that the most massive discs transform faster than lower mass discs.

% ------------------------------------------------------------------------------------------------------------

\section{Key factors affecting morphological transformation} \label{sec:keyFactors}

Section \ref{sec:clusterDiscs} shows that disc galaxies transform to non-discs at $z=0$ in larger fractions in clusters than in the field. The rate of morphological transformation of those galaxies that do become non-discy is also affected by the cluster environment, as signaled by the dependence on the time spent in the hosts. However, there is a significant amount of scatter in both the degree and timescale of transformation. Furthermore, we have seen that a fraction of the control discs also become less discy or non-discy by $z=0$, this fraction being larger for larger galaxy masses. This therefore implies that morphological transformations can also be induced outside of high-density environments and that, for the case of massive cluster galaxies, environment may act in addition to other, secular processes.

In this section, we explore what factors may affect either the degree or timescale of morphological change in order to understand the source of the galaxy-to-galaxy diversity and get insights on the underlying physical processes that may be responsible for the morphological changes. We consider not only several environmental/orbital parameters, but also intrinsic properties of the galaxies at accretion, merger histories, and AGN feedback and the impact each of these have on the total degree of change in morphology and the timescale to become non-discy $\Delta t_{\text{nondisc,last}}$.

\begin{figure*}
	\includegraphics[width=\linewidth]{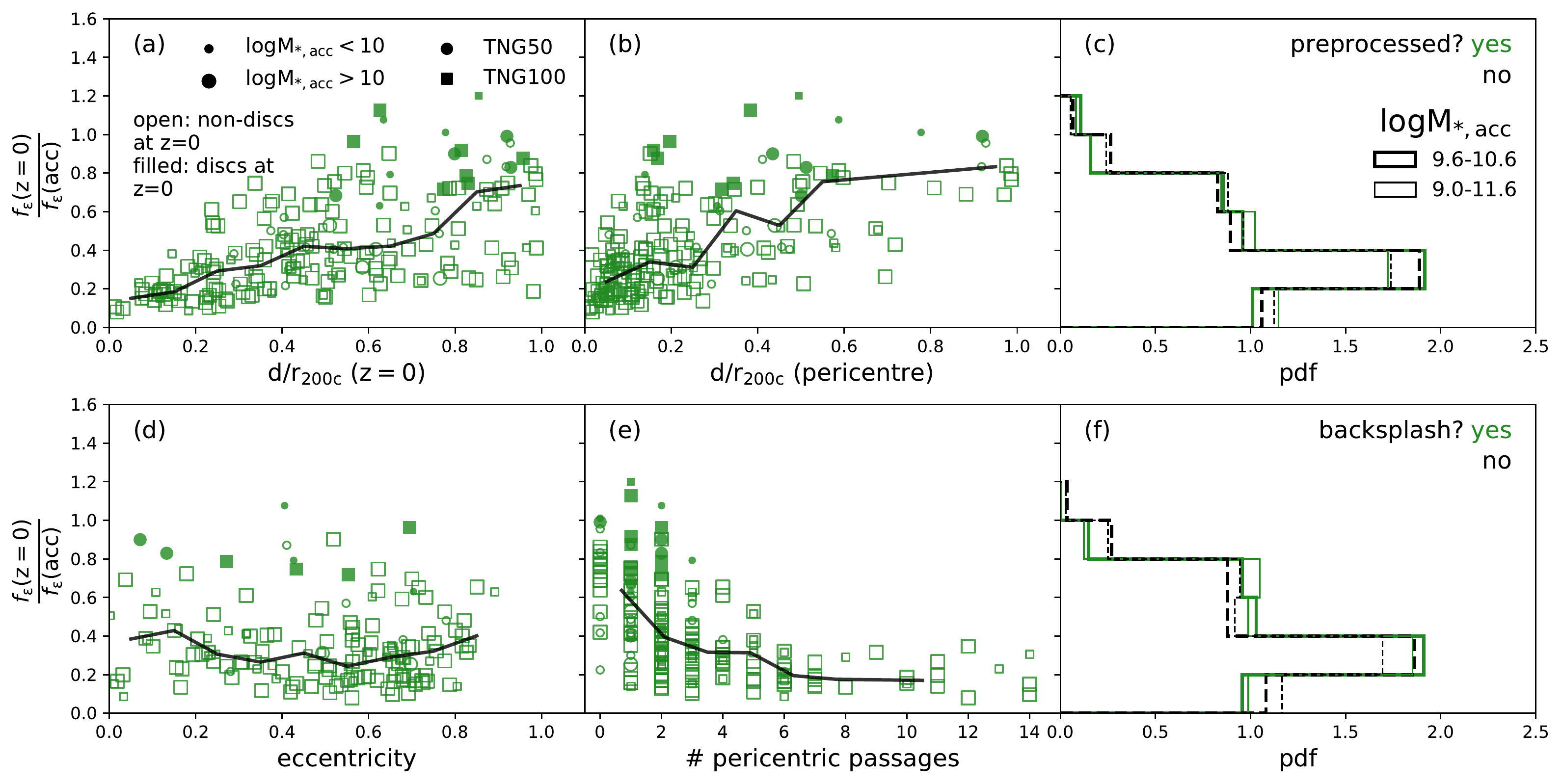}
	\caption{\textbf{Role of orbital properties.} Change in morphology from accretion to present day as a function of (a): present day radial distance normalized by cluster virial radius $d/r_{\text{200c}}(z=0)$, (b): pericentric/closest distance $d_{\text{peri}}//r_{\text{200c}}(z_{\text{pericentre}})$, (d): orbital eccentricity $e$, and (e): number of pericentric passages. The size of the symbol indicates the $M_{\text{*,acc}}$ of the galaxy. Galaxies that are still discs at $z=0$ are shown as filled symbols. Black lines show the running averages of $\mathit{f}_{\epsilon}(z=0)/\mathit{f}_{\epsilon}(\text{acc})$ in bins of the given orbital property. Note that galaxies where we cannot determine a pericentre (first infall) or apocentre (first outward portion of orbit) are excluded from panel (d). In panel (b), for galaxies on first infall, the values shown are the closest distance of the galaxies from the cluster. We also show the distribution of change in morphology separating cluster discs based on whether or not they are preprocessed (panel (c)) and whether or not they have had a backsplash event (panel (f)). Thick lines show the result for the two intermediate mass bins where we have both TNG50 and TNG100 data; thin lines show the full samples. Cluster discs that are found closer to the cluster, or have had closer or more numerous pericentric passages show the largest change to non-discy morphologies.} \label{fig:changeCircOrbital}
\end{figure*}

\subsection{The role of environmental factors} \label{sec:orbital}

Various environmental factors may contribute to differences in the degree and rate of morphological change between the cluster and control discs as well as the diversity within the cluster galaxies. We therefore examine the impact of the following orbital properties: present-day distance from the cluster centre normalized by the cluster's virial radius $d(z=0)/r_{\text{200c}}(z=0)$, the pericentric distance of the galaxy $d_{\text{peri}}//r_{\text{200c}}(z_{\text{pericentre}})$, orbital eccentricity $e$, number of pericentric passages, whether or not the galaxy has been preprocessed and whether it has had a backsplash phase. Note that while it is reasonable to expect that the host mass itself may affect the rate and timescales for morphological transformations, we omit such exploration as our host mass bin is relatively narrow (0.6 dex) and we have not seen any obvious trend with the 16 hosts at our disposal. We have confirmed that there is no significant dependence of the transformation timescales ($\Delta t_{\text{nondisc,last}}$, not shown) on any of these orbital parameters.

Fig. \ref{fig:changeCircOrbital} shows the change in morphology for the cluster discs as a function of the $d/r_{\text{200c}}(z=0)$, $d_{\text{peri}}//r_{\text{200c}}(z_{\text{pericentre}})$, eccentricity and number of pericentric passages (panels (a), (b), (d) and (e) respectively) and the distribution of $\mathit{f}_{\epsilon}(z=0)/\mathit{f}_{\epsilon}(\text{acc})$ for galaxies separated based on whether or not they were preprocessed (panel (c)) and whether they have had a backsplash event or not (panel (f)). The average change as a function of the first four properties is demonstrated by the running medians in bins of the given orbital property, shown by black lines in panels (a), (b), (d) and (e). Note that we have confirmed that the average trend is independent of the stellar mass of the galaxies at accretion. We discuss the results for each orbital property below.

\subsubsection{Present day distance to the cluster centre}
There is a clear correlation seen in Fig. \ref{fig:changeCircOrbital}(a), albeit with some scatter, between the degree of change in morphology and the present day location in the cluster: cluster discs found closest to the cluster centres exhibit the largest decrease in $\mathit{f}_{\epsilon}$ since accretion. This correlation may be connected to that with the time since accretion (top of Fig.~\ref{fig:times}), as galaxies with earlier accretion times tend to be found at smaller clustercentric distances.

\subsubsection{Pericentric distance}
Fig.~\ref{fig:changeCircOrbital}(b) also shows a clear correlation between the degree of change in morphology and the pericentric distance, although again with some scatter: closer pericentric passages are correlated with larger decreases in disciness since accretion. The pericentric distance is measured as the closest distance (normalized by the cluster $r_{200}$ at the time) the galaxy has ever been from the cluster centre. Note that galaxies on first infall are also considered here (7 and 13 cluster discs in TNG50 and TNG100 respectively) and their current position is taken as the pericentric distance.

\subsubsection{Orbital eccentricity}
We adopt a simple measure of orbital eccentricity $e$:
\begin{equation}
e = \frac{d_{\text{apo}}-d_{\text{peri}}}{d_{\text{apo}}+d{\text{peri}}}
\end{equation}
where $d_{\text{apo}}$ and $d_{\text{peri}}$ are the apocentric and pericentric distances. These are defined as the distances when the galaxy's orbit exhibits changes in radial direction; we do not interpolate between snapshots to compute them. $\sim10-15\%$ of the cluster discs in all three mass bins are on their first infalling orbit within their host clusters; a further $\sim12-16\%$ are on their first outgoing orbit, whereby we cannot define an apocentric distance for them respectively. These galaxies are excluded from analysis. For galaxies that have had multiple orbits within their host cluster, $e$ is measured for the first pericentric and apocentric passages.

Fig. \ref{fig:changeCircOrbital}(d) shows no strong dependence of the net change in morphology on orbital eccentricity. However, these findings are not necessarily in contrast with previous results that show that more radial orbits result in larger morphological changes, since we are unable to measure eccentricities for galaxies on first infall or on the first outward bound trajectory. Therefore it is possible that we do not see an underlying correlation between eccentricity and degree of morphological change, especially if most of these galaxies without eccentricity measurements are on radial orbits. Additionally, as we discuss later, galaxies on more radial orbits will spend less time near the central regions of the cluster; if this is an important factor in morphological transformation, this would also reduce any correlation between eccentricity and degree of change in morphology.

\subsubsection{Number of pericentric passages}
There is a strong correlation between the number of pericentric passages undergone by a cluster disc and its net change in $\mathit{f}_{\epsilon}$ since accretion as shown in  Fig.~\ref{fig:changeCircOrbital}(e) -- satellites that have had several pericentric passages show the largest changes in morphology towards non-discs, although there is significant scatter for galaxies that have had fewer pericentric passages. These results should be interpreted with caution however, since the number of pericentric passages is linked to the time spent in the cluster; however, the two correlations may be signposts of different physical mechanisms responsible for the morphological changes.

\begin{figure*}
	\includegraphics[width=\linewidth]{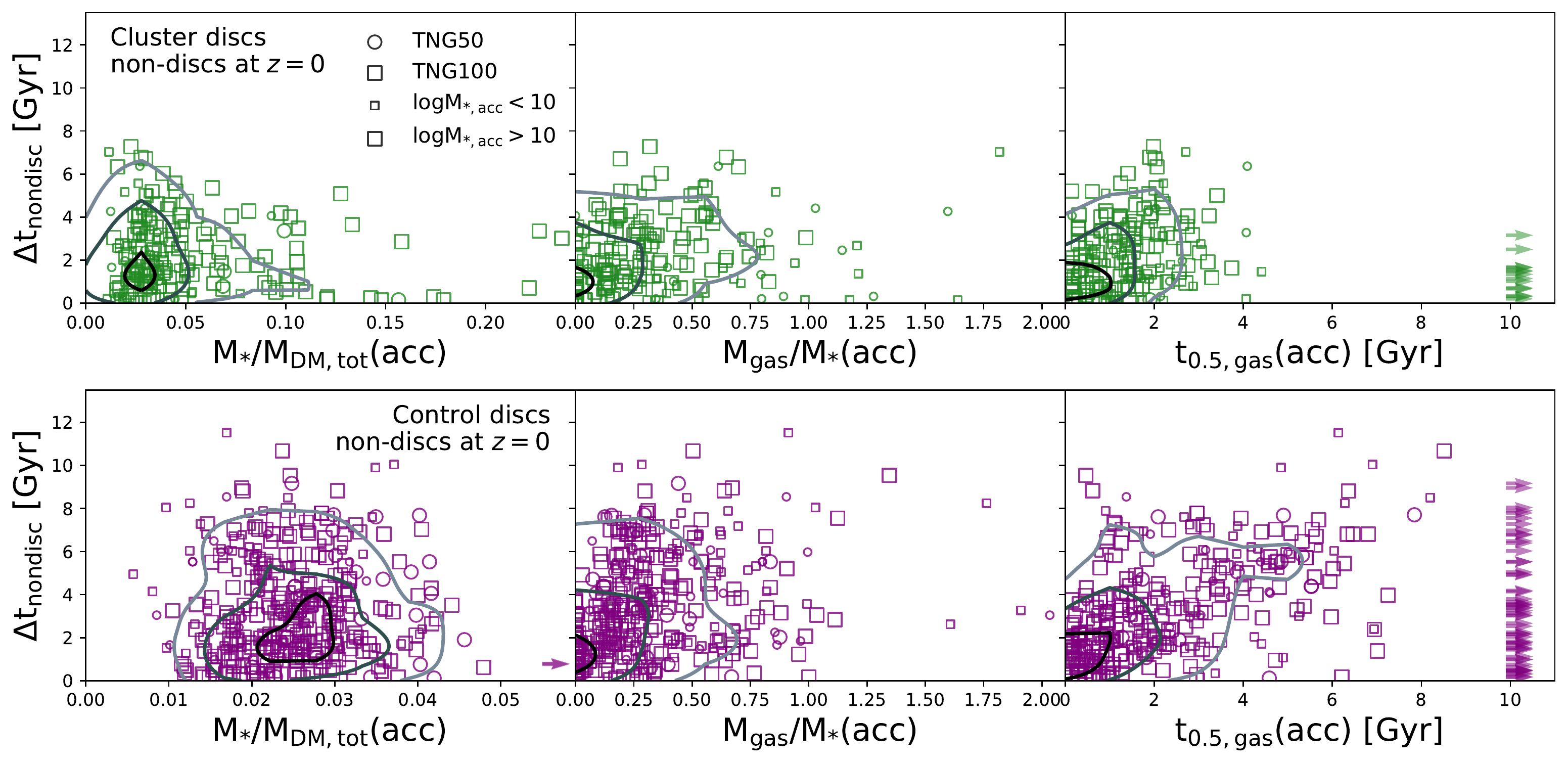}
	\caption{\textbf{Role of intrinsic properties.} Timescale to become non-discy $\Delta t_{\text{nondisc}}$ as a function of stellar to total DM mass $M_{*}/M_{\text{tot,bound}}$ at accretion (left), gas fraction $M_{\text{gas}}/M_{*}$ at accretion (middle), and timescale over which half of the initial (i.e. at accretion) gas mass is removed $t_{\text{0.5,gas}}$ (right) for the cluster (top) and control (bottom) discs. Only galaxies that become non-discs at $z=0$ are shown here. Note that $M_{\text{gas}}$ and $M_{*}$ are both measured within $2\times r_{\text{*,1/2}}$, whereas $M_{\text{tot,bound}}$ is the mass of all particles gravitationally bound to the galaxy. The size of the symbols denote the $M_{\text{*,acc}}$ of the galaxy. Black and grey lines show the median and 25$^{\rm th}$-75$^{\rm th}$ percentile ranges of $t_{\text{nondisc}}$ in bins of the given intrinsic property. Conditional 2D kernel density estimates are shown enclosing 20\%, 50\% and 80\% of the data. \emph{It is important to note that the cluster and control discs have significantly different ranges for their values of $M_{*}/M_{\text{tot,bound}}$ and $t_{\text{0.5,gas}}$.} The timescale of morphological change for both cluster and control discs shows a moderate dependence on the initial gas fractions of the discs and the rate at which their gas is subsequently depleted. The trends are stronger for more massive galaxies.} \label{fig:avgEvolCircIntrinsic}
\end{figure*}

\subsubsection{Preprocessing}
As mentioned earlier, $\sim52-66\%$ of the cluster discs have been preprocessed in smaller groups before accretion onto their final host, and therefore, are likely to have been affected by these group environments as well. However, we find no significant difference in the degree of morphological change of galaxies that were preprocessed and those that were not (Fig. \ref{fig:changeCircOrbital}(c)). This does not mean that preprocessing is not important in general, as in fact it does determine the very morphological mix of galaxies at accretion into their $z=0$ hosts (as discussed in Section~\ref{sec:sampleZ0}).

\subsubsection{Backsplash vs. non-backsplash}
Finally, backsplashing galaxies may temporarily avoid the influence of a dense cluster environment, which could lead to slower overall rates of morphological evolution. We define backsplash galaxies as those that have been outside the virial radius of the host cluster for any amount of time after accretion and compare them to non-backsplash galaxies that remain within $r_{200c}$ after accretion. From Fig. \ref{fig:changeCircOrbital}(f), it is evident that both backsplash and non-backsplash galaxies have nearly identical distributions of net change in morphology.

\begin{figure*}
    \includegraphics[width=\linewidth]{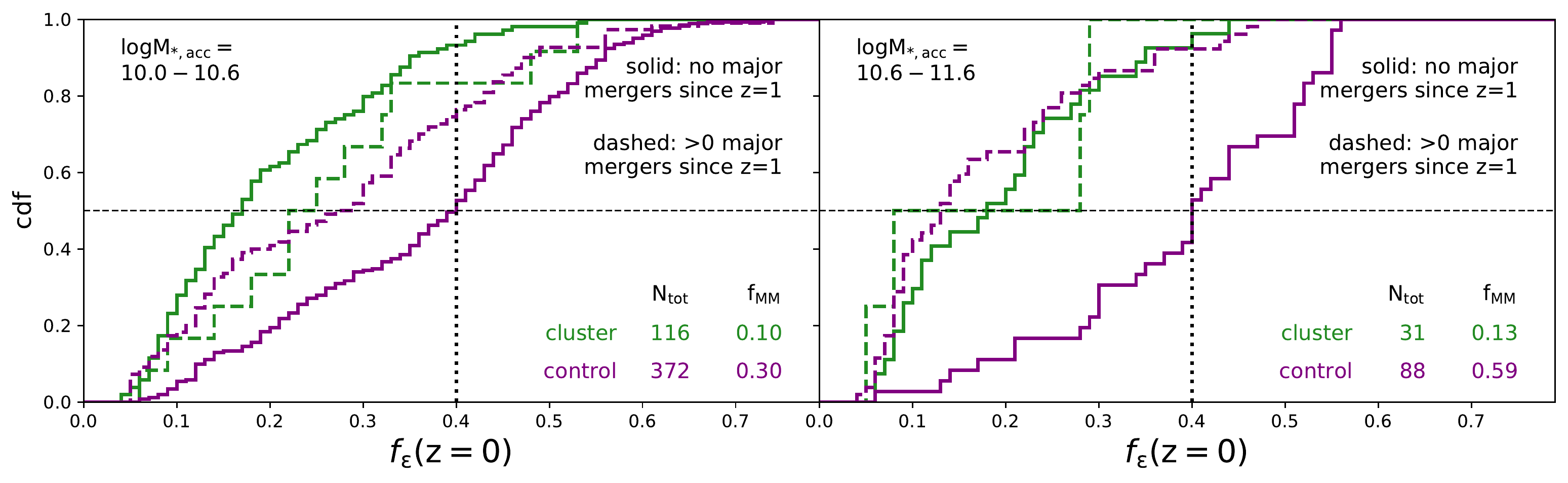}
    \caption{\textbf{Role of mergers.} Distribution of $z=0$ morphology for cluster and control discs separated based on whether they have had any major mergers since $z=1$. Major mergers are defined as those with mass ratios greater than 1:4. Galaxies with no major mergers are shown by the solid lines, those with at least one major merger by dashed lines. $N_{\text{tot}}$ provides the total number of galaxies in each subsample and $f_{\text{MM}}$ the fraction of the subsample that have had a major merger. The two panels show the results for the two higher bins of stellar mass at accretion, including data from the TNG50 and TNG100 samples. The horizontal dashed line indicates the median of the subsamples, the vertical dashed line indicates the separation between discs and non-discs. The occurrence of a merger in a galaxy's history significantly affects its final morphology only in the case of the most massive control discs.} \label{fig:mergerEffect}
\end{figure*}

\subsection{The role of intrinsic properties} \label{sec:intrinsic}

In addition to environmental effects, certain intrinsic properties of the galaxies may also influence, or even determine, their morphological evolution. In particular, we consider the following galaxy properties: gas fraction i.e. gas-to-stellar mass ratio at the time of accretion, the ratio of stellar mass to \emph{total} dark matter mass at accretion, and the timescale over which their gas is depleted after accretion. In Fig. \ref{fig:avgEvolCircIntrinsic}, we show the timescale to become non-discy $\Delta t_{\text{nondisc,last}}$ (as defined in Section~\ref{sec:methodsTimescales}) for the cluster (top panels) and control (bottom panels) discs as a function of these three intrinsic properties. We have also examined the net change in $\mathit{f}_{\epsilon}$ since accretion and find no significant dependence on any of these intrinsic properties, except for a few cases highlighted below.

\subsubsection{Stellar to total DM mass ratio at accretion}
We first consider the stellar-to-total DM mass ratio of the galaxies at accretion, to determine whether the presence of a more massive DM halo can affect the stellar morphological transformation. The left panels of Fig.~\ref{fig:avgEvolCircIntrinsic} show $\Delta t_{\text{nondisc,last}}$ as a function of the initial stellar-to-total DM mass ratio $M_{*}/M_{\text{DM,tot}}(\text{acc})$. Note that here $M_{\text{DM,tot}}$ is the mass of all gravitationally bound DM particles of the subhalo. Importantly, the cluster and control discs span different ranges in initial stellar-to-DM mass ratios, a further manifestation of the preprocessing of the cluster discs, which have likely lost some dark matter in their previous host haloes \citep[see also][]{Joshi19,Engler20}; even without actual preprocessing, the galaxies have likely experienced the effects of their future hosts at large distances, which would also contribute to the reduction in initial stellar-to-DM mass ratios \citep[e.g. see][]{Behroozi14}. 

There is no significant correlation between the initial $M_{*}/M_{\text{DM,tot}}(\text{acc})$ and the time for the galaxies to become non-discs, especially for the control discs; for the cluster discs with high ($\gtrsim0.1$) values of $M_{*}/M_{\text{DM,tot}}(\text{acc})$, there is a marginal preference for shorter transformation timescales. Thus cluster discs with high initial $M_{*}/M_{\text{DM,tot}}(\text{acc})$ evolve marginally faster towards non-discs.

\subsubsection{Gas fraction at accretion}
\label{sec:gasFraction}
Some studies have shown that the gas fraction of galaxies can affect their resilience to morphological transformations during mergers, with gas-rich discs either retaining more of their discs or re-establishing discs after the merger \citep[e.g.][]{Springel05,Robertson06,Hopkins09,Kannan15,Martin18}. Similarly, the presence of a significant gas reservoir in the galaxies may also impart resilience against morphological transformations due to the group or cluster potential. We therefore study the effect of gas fraction at accretion $M_{\text{gas}}(\text{acc})/M_{*}(\text{acc})$ on morphological transformation.

The middle column of Fig.~\ref{fig:avgEvolCircIntrinsic} shows $\Delta t_{\text{nondisc,last}}$ as a function of the initial gas fraction $M_{\text{gas}}/M_{*}(\text{acc})$. For the few galaxies that have zero gas mass at accretion, note that we have used their last non-zero gas mass before accretion. Unlike the stellar-to-DM mass ratios, the cluster and control discs have similar distributions of initial gas fractions. We emphasize here that the gas mass shown is the sum of the mass of all gravitationally bound gas cells within twice the \emph{stellar} half mass radius. For most galaxies, especially in the control sample, there is considerable gas mass beyond this aperture and the initial gas fractions are therefore significantly dependent on the aperture used. Nonetheless, for any given operational definition of gas mass and across all stellar mass ranges, we find a weak (if any) correlation between the initial gas fraction and the timescale of morphological transformation for both cluster or control discs. We also find a weak dependence on the gas fraction at accretion for the degree of morphological change (not shown for brevity). Exceptions to these statements seem to appear at the highest-mass end (initial stellar mass above a few $10^{10}\MSUN$) -- both in clusters and in the field, the more (initially) gas-rich galaxies exhibit relatively smaller changes in morphology and longer timescales for the change than the more gas-poor ones.

\subsubsection{Rate of gas removal}
\label{sec:rateGasRemoval}
Finally, we consider the rate at which the discs lose their gas content after accretion, which is indicative of their rate of quenching. To quantify the rate of gas removal, we calculate $t_{\text{0.5,gas}}$, the time intervals between accretion and when the galaxy has lost 50\% of its gas mass at accretion. The right panels of Fig.~\ref{fig:avgEvolCircIntrinsic} show $\Delta t_{\text{nondisc,last}}$ as a function of $t_{\text{0.5,gas}}$. Galaxies that are yet to lose more than half their initial gas mass are assigned an arbitrarily high value of $t_{\text{0.5,gas}}$ and shown as arrows. Importantly, note that the range of values of $t_{\text{0.5,gas}}$ is significantly different for the cluster and control discs. 

While there is no strong correlation seen between the gas depletion timescale and morphological transformation timescale for the cluster discs, we do see a significant correlation for the control discs, such that control discs that retain their gas for longer also transform their morphology on longer timescales. The cluster environment appears to remove any such correlation for the cluster discs, possibly by simply drastically reducing the range of possible timescales for gas depletion, from up to 6-8 Gyr for the control galaxies down to 3-4 Gyr at most for cluster galaxies. This is consistent with previous findings with IllustrisTNG galaxies, where it has been shown that ram-pressure stripping acts rapidly at removing gas for satellite galaxies \citep{Yun19}.

\begin{figure*}
    \includegraphics[width=\linewidth]{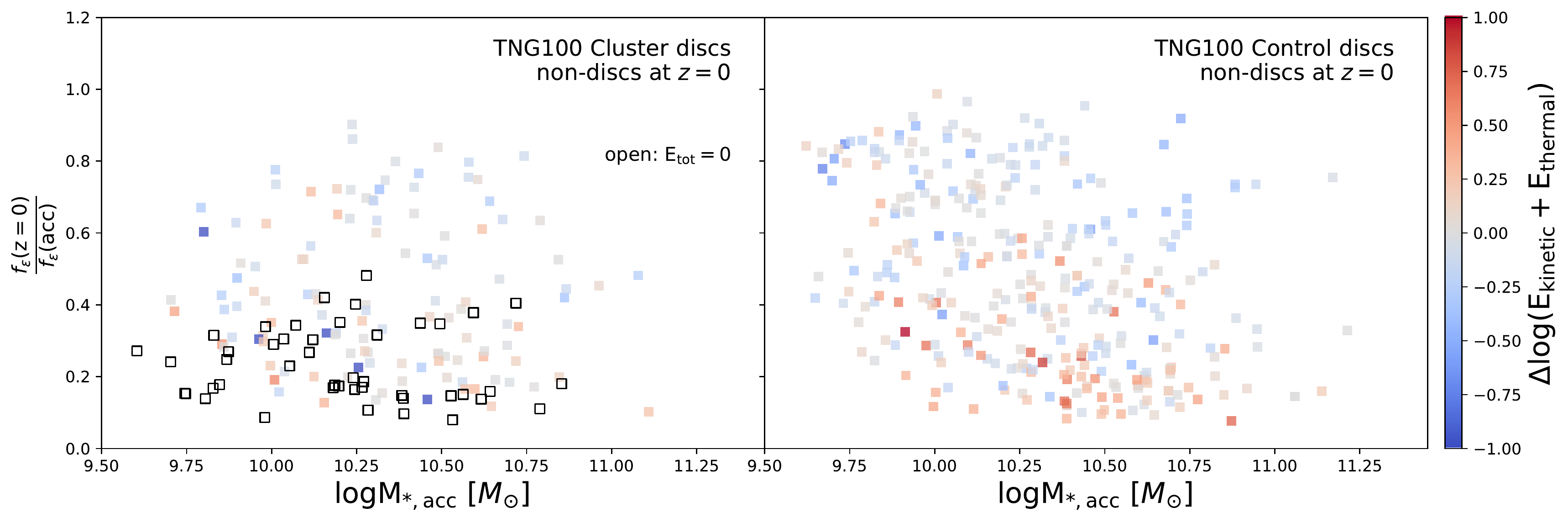}
    \caption{\textbf{Role of AGN feedback.} Change in morphology as a function of stellar mass at accretion, showing the dependence on total energy injected through AGN feedback over the lifetime of the galaxy, for the TNG100 samples. Only cluster and control discs that are non-discs at $z=0$ are shown here. The colour of each datapoint represents the difference in $\log{E_{\text{tot}}}$ of the galaxy and the median value in 0.2 dex bins of stellar mass at accretion. A small number of the cluster discs have no black holes; these are shown by the open black symbols. At fixed stellar mass, control discs with $M_{\text{*,acc}}\gtrsim10^{10}\MSUN$ that have had relatively higher amounts of total BH feedback energy are less discy. A similar but weaker correlation is also present for cluster discs.} \label{fig:AGNeffect}
\end{figure*}

\begin{figure}
    \includegraphics[width=\linewidth]{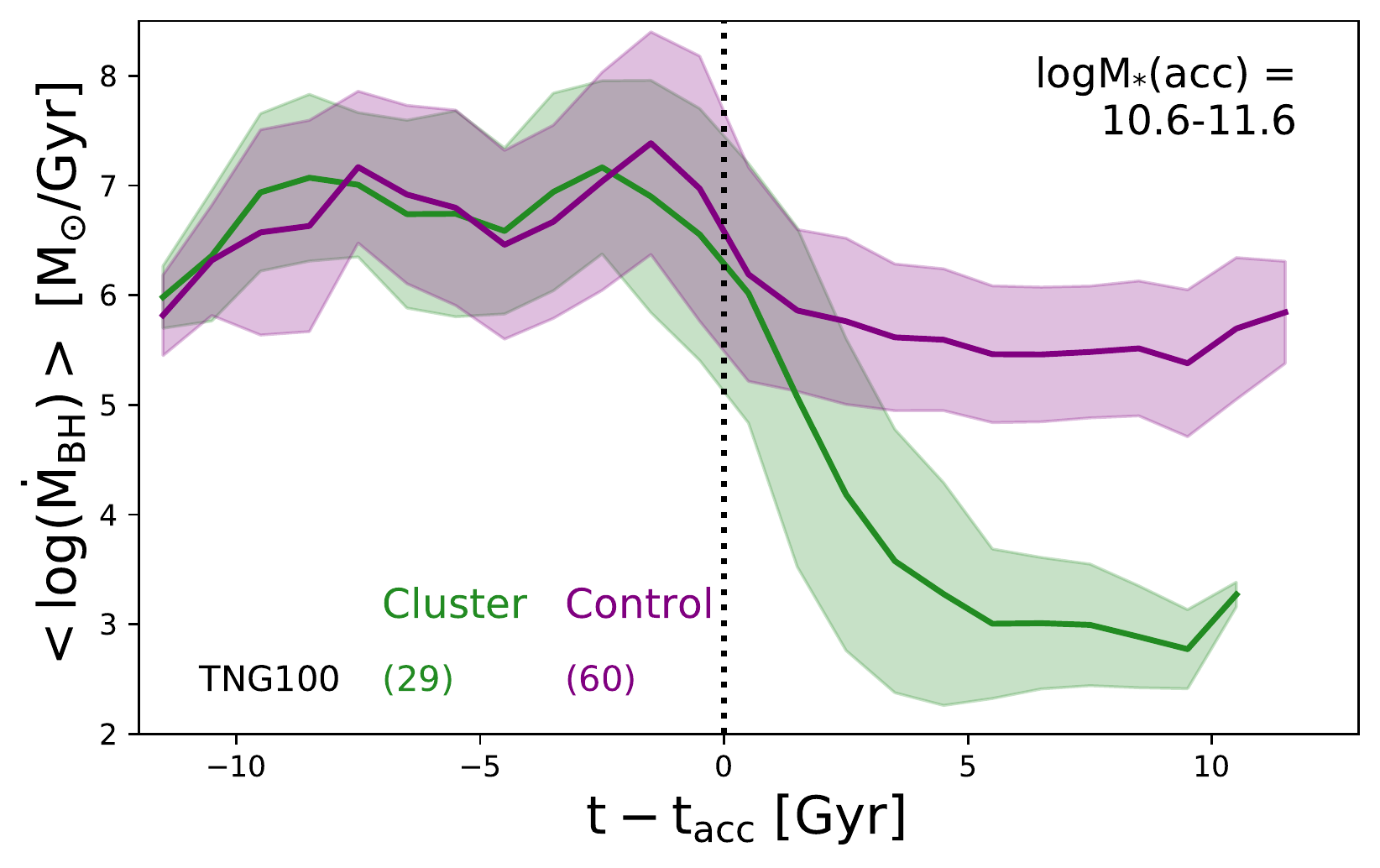}
    \caption{Average of the instantaneous BH accretion rates $\log{\dot{M}_{\text{BH}}}$ as a function of time relative to accretion for the most massive bin of galaxies. Values shown are the average in 0.5 Gyr bins of time relative to accretion. Only galaxies that are non-discs at $z=0$ are included here. Shaded regions show the $1\sigma$ galaxy-to-galaxy variation. BH accretion is significantly suppressed for the cluster discs after accretion. There is similar suppression of BH accretion in the lower mass bins of galaxies, both for TNG50 and TNG100; in fact, the differences between the cluster and control discs are larger for less massive galaxies.} \label{fig:avgBHAccretion}
\end{figure}

\subsection{The role of mergers} \label{sec:mergers}

In Fig.~\ref{fig:mergerEffect}, we investigate the effect that the occurrence of mergers in the galaxies' histories has on their morphological transformation. We show the cumulative distribution functions of the cluster and control disc samples, separating them into those that have had no major mergers since $z=1$ and those that have had at least one. Major mergers are defined as those with stellar mass ratios of $>1/4$. The subsamples are further divided into bins of stellar mass at accretion. We only show the results for the two most massive bins; the results for the $\log{M_{\text{*,acc}}}=9.6-10$ bin are nearly identical to those for the $\log{M_{\text{*,acc}}}=10-10.6$ bin, while we do not have enough statistics in the lowest mass bin to make a conclusive statement. We have combined the TNG50 and TNG100 samples for better statistics, but it should be noted that the most massive bin of $\log{M_{\text{*,acc}}}=10.6-11.6$ only contains TNG100 data. Furthermore, we do not distinguish between mergers that occur before or after accretion for the cluster discs, even though they are more likely to occur before accretion. In the two mass bins depicted in Fig.~\ref{fig:mergerEffect}, over half the cluster galaxies have had no major mergers since $z=1$; this is also the case for the control galaxies in the $\log{M_{\text{*,acc}}}=10.0-10.6$ mass range.

Fig. \ref{fig:mergerEffect} shows that the occurrence of one or more major mergers since $z=1$ does not significantly affect the final morphologies of the cluster disc samples of any mass. This is not the case for the control discs -- galaxies that have had at least one major merger since $z=1$ are more likely to be non-discy at $z=0$ than those that have had no major mergers. The effect is largest for the most massive galaxies. Similar conclusions hold if we also include minor mergers since $z=1$ as well as if we consider mergers since $z=2$.

The fact that in the most massive bin of control discs, over half the galaxies have had major mergers and that this is correlated with relatively more galaxies becoming non-discy by $z=0$ is consistent with what was seen in Figs. \ref{fig:circVsMstar} and \ref{fig:discDestructionFrac}, where a higher proportion of high-mass control discs are non-discs at $z=0$ compared to low mass ones. Most galaxies that have had major mergers since $z=1$ have experienced either one or two such events. We have confirmed that the \emph{number} of major mergers has no significant effect on the galaxies' morphological transformation, only whether or not they have experienced one. Thus, it is clear that mergers play an important role in the morphological transformation of the control discs, especially for more massive galaxies, but not for the cluster discs.

\subsection{The role of AGN feedback} \label{sec:agn}

The results so far show that galaxies at the higher-mass end ($\gtrsim10^{10}\MSUN$) display different trends compared to lower-mass galaxies: in the field, they exhibit larger losses of disciness and overall change their stellar morphology much more rapidly than their lower-mass analogues. Additionally, high-mass control discs change their morphology from discs to non-discs somewhat more rapidly than even cluster discs of similar mass -- although the median times to become non-discs are similar, about 1.6 and 2 Gyr for massive control and cluster galaxies respectively, a larger number of massive control discs become non-discs faster in comparison to cluster discs (see Fig.\ref{fig:times}).

Our two highest-mass bins ($10^{10-10.6}$ and $10^{10.6-11.6}\MSUN$) correspond roughly to the mass where AGN feedback begins to show an effect on the SF activity of galaxies, at least in the IllustrisTNG model \citep{TNGNelson18,Donnari19,Terrazas20,Zinger20}. For example, the fraction of quenched central galaxies increases from the percent level to more than half of the galaxy population precisely around the mass scale of a few $10^{10}\MSUN$, both in observations and simulations. One possible scenario that may explain our findings is if the AGN activity plays a role in determining the transformation of the control discs to non-discs. In fact, it has been demonstrated that, at least within the IllustrisTNG framework, it is the BH feedback in its kinetic mode that establishes the quenched population of high-mass galaxies \citep{TNGMethodsWeinberger17, TNGNelson18, Terrazas20}. Should the cluster environment suppress this AGN activity, this could lead to the most massive cluster discs somewhat slowing down their rate of morphological change from what they would otherwise undergo outside of a dense environment.

In Fig. \ref{fig:AGNeffect}, we show the change in morphology from accretion to $z=0$ as a function of stellar mass at accretion for the cluster and control discs that become non-discs by $z=0$. The colour of the data points indicates the difference between the log of the total energy injected by the AGN feedback over the lifetime of the galaxy and the median value in 0.2 dex bins of $M_{\text{*,acc}}$. As the energy injected by AGN feedback is known to be strongly correlated with stellar mass, examining the difference to the median value allows us to determine if there is a residual effect of AGN feedback on the degree of morphological change of the galaxies. We measure the energy as the sum of the energy injected by all black holes found at $z=0$ in the galaxy, at each snapshot, including both the kinetic and thermal feedback modes. Note that a few cluster discs have zero total energy injected from a black hole, and are shown as open black symbols. This occurs because the galaxies have no BH at $z=0$, largely due to numerical effects. In the IllustrisTNG model, BHs are seeded at the FOF-halo finding step, at the centres of haloes above a given threshold mass that do not already have a BH \citep[see][for further details]{TNGMethodsWeinberger17}. The BH is repositioned to the local minimum in potential at each large timestep. Therefore, if a subhalo was accreted early enough, it may never have had a BH. Additionally, a subhalo may `lose' its BH during a close pericentric passage, where the BH may be repositioned to the central galaxy in the halo, somewhat anticipating what may occur as a BH-BH merger event at a later epoch.

Fig. \ref{fig:AGNeffect} shows that the integrated energy injected via AGN feedback does seem to correlate with the morphology of control discs with $\log{M_{\text{*,acc}}}\gtrsim10$ -- control galaxies with relatively higher AGN energy injection exhibit a higher change in $\mathit{f}_{\epsilon}$. This correlation is also seen for the cluster discs in the same mass range, although the dependence is milder. Note that the difference in the median BH feedback energy for the cluster and control discs is approximated 0.15-0.4 dex; however, this is the total energy injected over the lifetime of the galaxy. As we show next, there is significant reduction in BH feedback for the cluster discs after accretion. There is no significant correlation seen for galaxies with $\log{M_{\text{*,acc}}}\lesssim10$ for either cluster or control discs.

The milder correlation between morphological changes and AGN feedback in clusters than in the field may in fact be due to two possibilities: either the environmental effects driving the morphological transformation are so strong that they cancel out more subtle dependencies and/or the environment affects the very activity of the BHs, in turn changing their impact on morphological changes. The latter possibility is quantified in Fig. \ref{fig:avgBHAccretion}, which demonstrates that the most massive cluster discs have significantly suppressed BH accretion rates after accretion onto the cluster; we have confirmed that this is the case in all mass bins -- in fact, the effect is strongest for lower mass galaxies. The suppression of BH accretion rates for the cluster discs is a direct result of the little gas content in these galaxies, as BH particles accreted from nearby gas cells at the Bondi-Hoyle-Lyttleton accretion rate in the IllustisTNG model \citep{TNGMethodsWeinberger17}. Our findings appear to be in contrast to previous studies that have concluded that higher density cluster environments actually trigger enhanced BH activity in satellites for a period of time, which in turn acts to eventually quench the satellites \citep[e.g. see][]{McGee13,Poggianti17,Ricarte20}. However, it should be kept in mind that the results shown in Fig. \ref{fig:avgBHAccretion} include a specific selection of galaxies, those that were discs at accretion and transform to non-discs by $z=0$, and therefore they represent a different subset of the galaxy population than those of the previous studies. Additionally, Fig. \ref{fig:avgBHAccretion} averages over $\log{\dot{M}_{\text{BH}}}$ for several galaxies and over 0.5 Gyr timescales and therefore necessarily does not show short-lived bursts of AGN activity. It is beyond the scope of this paper to investigate environmental effects on AGN activity for all cluster satellites.

Figs.~\ref{fig:AGNeffect} and \ref{fig:avgBHAccretion} indicate that AGN feedback may play an active role in determining the net change in morphology for more massive control galaxies and that the cluster environment lessens this effect by significantly suppressing BH accretion rates for the cluster discs. In fact, we have checked that the correlation at fixed mass between morphology and BH injected energy is in place at the high-mass end even when thermal and kinetic feedback modes are considered separately \citep[although it disappears when the ratio of kinetic to thermal energy injection is considered, in contrast to what occurs for galaxy colors and SFRs,][]{TNGNelson18, Terrazas20}. Furthermore, we have confirmed that the trends seen in Fig.~\ref{fig:AGNeffect} are maintained when we only consider galaxies that have had at least one major or minor merger since $z=1$, but are largely removed when only considering galaxies that have had no mergers since $z=1$. This indicates that the role of increased AGN feedback may be to support any further morphological changes that galaxies may undergo especially at the high mass end. In the case of the most massive cluster discs, by suppressing BH accretion, the cluster environment has two opposing effects -- directly causing morphological transformation through interaction with the cluster potential and reducing the impact of the transformation through reduced AGN feedback. As a final remark, it should however be noted that the connection we see in Fig.~\ref{fig:AGNeffect} could be an indirect manifestation of the correlation between mergers and BH activity, a point we return to in Section \ref{sec:discussion} and a topic of great debate in the literature \citep[e.g. see][for one of the latest assessments of this connection from the perspective of the EAGLE simulations]{McAlpine20}.

\section{Discussion and Interpretations} \label{sec:discussion}

\begin{figure*}
    \includegraphics[width=\linewidth]{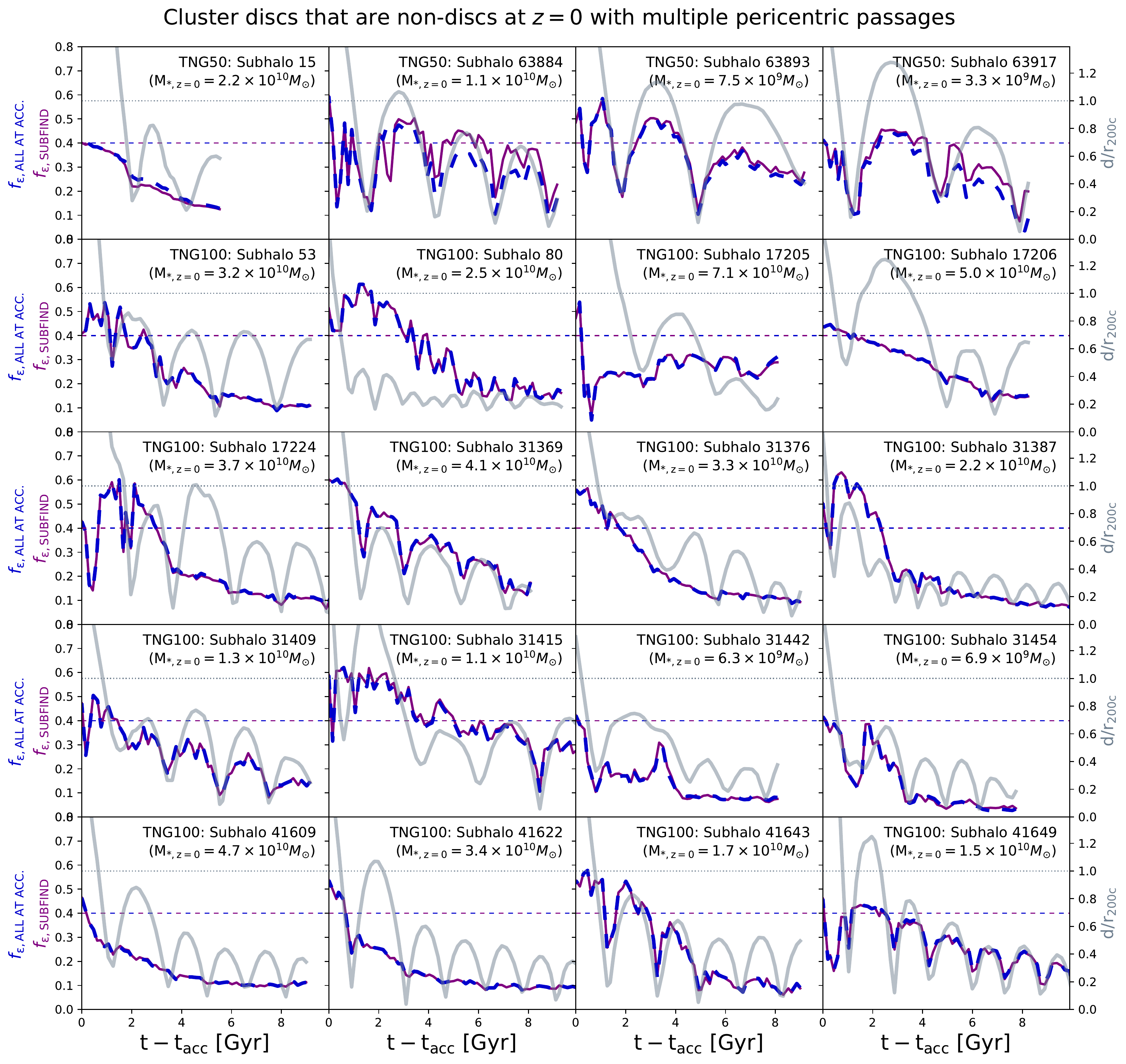}
    \caption{{\textbf{Impulsive changes of morphology at pericentric passages.} Each panel represents the orbital evolution of a cluster galaxy, either from TNG50 or TNG100, as indicated. A random selection of galaxies with several pericentre passages are shown. Gray curves show the distance of the galaxy to the host cluster, normalized by the cluster's virial radius at the time; dashed blue and solid magenta curves represent the evolution of the galaxy's circularity fraction $\mathit{f}_{\epsilon}$ measured in two different ways to check for robustness (see text for details). The two morphology measures are consistent with each other and show that cluster discs undergo dramatic morphological changes upon passing close to the centre of their host potential, sometimes even re-establish more discy configurations as they orbit outwards, and yet see an overall gradual decrease of disciness, both in the minima and maxima of their $\mathit{f}_{\epsilon}$ time evolution.}}
\label{fig:CircAtpericentre}
\end{figure*}

\subsection{Phenomenology of the transformations in clusters}

\begin{figure*}
    \includegraphics[width=\linewidth]{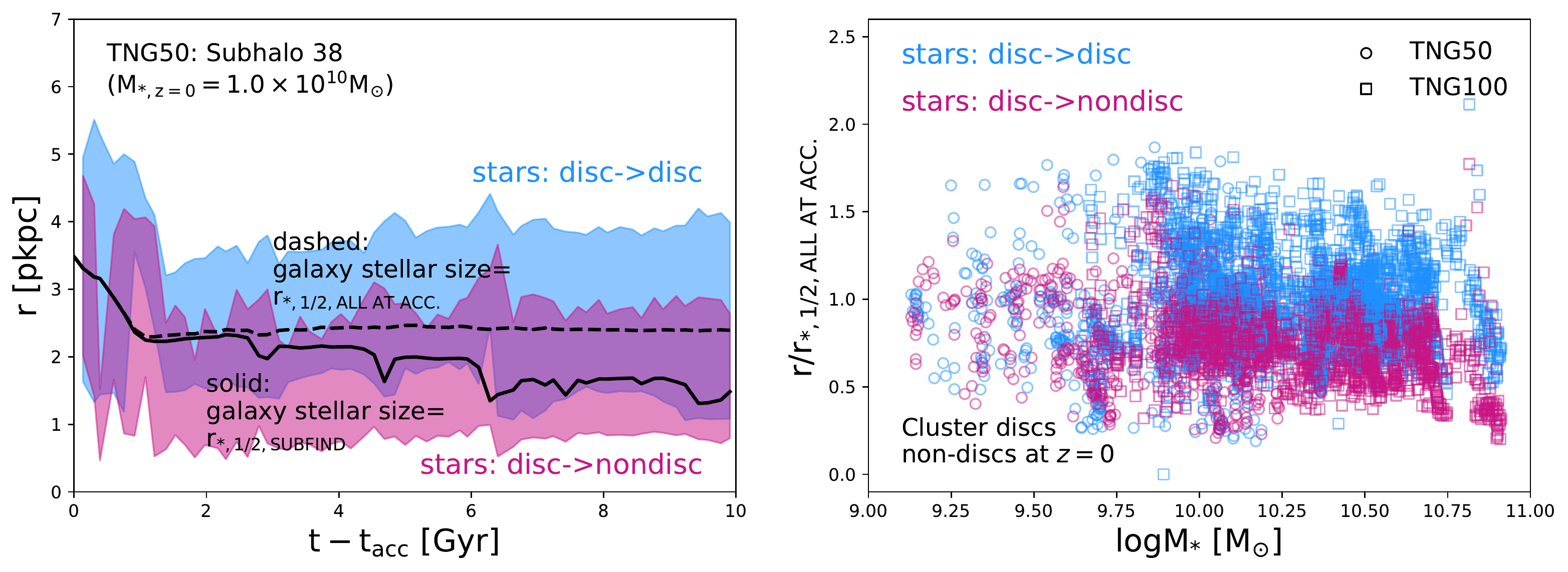}
    \caption{\textbf{Heating of centrally-concentrated stars.} \emph{Left:} Initial radial distribution of disc stellar particles that remain disc stars (blue) and disc stars that become non-disc stars (pink) in the following snapshot, as a function of time since accretion. Shaded regions show the $20^{\rm th}-80^{\rm th}$ percentile of the initial radial distribution. The half-mass radius measured using \textsc{subfind} particles and using all particles bound to the galaxy at accretion are also shown by the black solid and dashed lines respectively. \emph{Right:} Median radius of disc stars that remain disc stars (blue) and disc stars that become non-disc stars (pink) as a function of galaxy stellar mass. Each point represents a single snapshot for one galaxy. We show all TNG50 cluster discs that are non-discs at $z=0$ and all TNG100 cluster discs that are non-discs at $z=0$ and that have had multiple pericentric passages.}
\label{fig:heating}
\end{figure*}

\subsubsection{Impulsive processes near the cluster centres}
As seen in Fig.~\ref{fig:changeCircOrbital}, the degree of morphological transformation of the cluster discs is dependent on the pericentric distances of their orbits and on the number of pericentric passages: closer and more numerous pericentric passages correlate with larger degrees of morphological changes towards less discy and non-disc configurations. We have checked that, even for similar accretion times, e.g. between 8-12 Gyr ago, galaxies that underwent a larger number of passages close to the centre of their host potential still exhibit higher losses of disciness than analogue galaxies with fewer pericentric passages. Conversely, for galaxies with the same number of pericentric passages, the degree of morphological change shows little dependence on accretion time, indicating that the amount of time spent in a cluster since accretion is a secondary driver. Larger numbers of pericentric passages within similar intervals of time would translate to shorter time intervals actually spent at pericentre i.e. near the cluster. This may indicate that the property of interest for morphological change is the time interval near pericentre, rather than the total time spent within the cluster. Furthermore, from a population perspective, it is unlikely (about 10 per cent) for cluster discs that have at least one pericentric passage to still be discy at $z=0$, this being the case across the range of satellite stellar mass and accretion times.

In Fig.~\ref{fig:CircAtpericentre}, the orbital and time evolution of a random selection of cluster disc galaxies with multiple pericentric passages are shown for both TNG50 and TNG100. Gray curves denote the distance of the galaxy to the host cluster center, normalized by the cluster's virial radius at the time. Solid magenta curves represent the evolution of the galaxy's circularity fraction $\mathit{f}_{\epsilon}$ measured as defined in Section~\ref{sec:methodsMorph}, i.e. within twice the stellar half-mass radius and accounting for the stellar particles that are gravitationally bound to the satellite according to the \textsc{subfind} algorithm. Since \textsc{subfind} may encounter difficulties in separating particles belonging to a subhalo (i.e. satellite) from those belonging to the cluster's central galaxy in the high-density regions at the centres of haloes, we test the measurement of disciness adopted thus far by re-measuring the circularity fraction accounting for all the stellar particles a satellite has at accretion; more details are given in Appendix~\ref{app:checksOnSubfind} and results are shown with dotted blue curves in Fig.~\ref{fig:changeCircOrbital}. The two morphology measures are consistent with each other demonstrating that our morphological estimators are robust against the subtleties of the unbinding procedure in the adopted subhalo finder. 

Fig.~\ref{fig:CircAtpericentre} (supported by the tests in Appendix~\ref{app:checksOnSubfind}) demonstrates that the cluster discs undergo drastic changes in $\mathit{f}_{\epsilon}$ during pericentric passages by becoming less discy during infall and then often \emph{reestablishing} their discs during the outward portions of their trajectories. There is however a gradual reduction in $\mathit{f}_{\epsilon}$ along with these cyclic changes associated with the galaxies' orbits -- depending on the galaxy under scrutiny, a decline of $\mathit{f}_{\epsilon}$ can be seen in both the local minima at pericentre and at the local maxima at apocentre. We have investigated the exact method by which discs may be reestablished -- in general, the disc is reformed by perturbed stars resettling into disc orbits on short timescales i.e. between consecutive snapshots. However, on long timescales, continued star-formation is likely to be an important factor in the rejuvenation of the discs as discussed in Section \ref{sec:discussionSFRvsFcirc}. These results suggest that the morphological transformation is a result of an impulsive process near the cluster centre.

\subsubsection{Heating of disc stars}
The morphological transformation of cluster discs, which we have shown is consistent with being triggered by an impulsive physical phenomenon acting on short timescales, can proceed in two potential ways: (i) by tidally stripping away the outer disc material from the galaxy while leaving the more dispersion-dominated non-discy inner regions intact or (ii) by altering the orbits of the stars such that they are no longer confined to a disc.

To understand which of these two channels is dominant or if they each play equally significant roles, we examine how the stellar particles in our simulated galaxies are affected when galaxies change their level of disciness. For each galaxy at any given time, we can identify four classes of stellar particles: disc and non-disc stellar particles that either remain so in the next snapshot or convert to non-disc and disc particles, respectively. As stated in Section \ref{sec:methodsMorph}, we define stellar particles to be `disc' particles if they have a circularity $\epsilon>0.7$, `non-disc' if $\epsilon \leq 0.7$. To circumvent possible issues with the unbinding procedure of \textsc{subfind}, for each galaxy we consider all stellar particles that were bound to the galaxy at accretion and that at any given time are found within twice the stellar half-mass radius of all such stars at accretion, $r_{\text{*,1/2,ALL~AT~ACC.}}$. 

The left panel of Fig.~\ref{fig:heating} shows the evolutionary track of an example galaxy; the solid black line denotes the time evolution, after accretion, of the galaxy stellar size (i.e. half mass radius) according to \textsc{subfind} and the dashed black the time evolution of the stellar half-mass radius including all stars that were gravitationally bound at accretion. All radii are in physical kpc. The shaded regions show the $20^{\rm th}-80^{\rm th}$ percentile distributions of the galactocentric distances of two categories of stellar particles in a given snapshot: stars that are on disc orbits that remain on disc orbits in the next snapshot (blue) and stars that change from a discy orbit to a non-discy orbit in the next snapshot (magenta). In this galaxy, former disc particles that remain disc particles are preferentially found farther away from the galaxy centre in comparison to former disc particles that become non-disc particles. In other words, stars that move from disc to non-disc orbits are relatively more centrally-concentrated than those that remain discy.

The situation exemplified by this one TNG50 galaxy is confirmed on average across all galaxies. In the right panel of Fig.~\ref{fig:heating}, we consider all TNG50 and TNG100 cluster galaxies that are discs at accretion and become non-discy by $z=0$ (for TNG100, only galaxies that have had more than 3 pericentric passages are shown for clarity). For each of them, we investigate all available snapshots in time along their orbit and at each time consider the galaxy's stellar particles similarly as above: those that were on disc orbits and remain so in the next snapshot (blue) and those that transition to non-disc orbits (magenta). Data points represent the medians of the galactocentric distances of the two categories of stars normalized by their galaxy stellar size at the time of observation, each galaxy appearing multiple times in the plot depending on its accretion time. Across all galaxies and on average, stars that move from disc to non-disc orbits are relatively more centrally-concentrated than those that remain disc-like. Note that this could be an indicator for the formation/growth of a bulge rather than an overall morphological transformation. However, we have confirmed that there is significant overlap in the radial distributions of both sets of particles, with a non-negligible portion of particles that remain on disc orbits (move to non-disc orbits) being found at $\sim 0.5\,r_{*,1/2}$ ($\sim 1.5\,r_{*,1/2}$). Hence, although bulge formation is likely an important contributor to the change in $\mathit{f}_{\epsilon}$, it is not the only process of morphological transformation of the cluster discs. Importantly, the disc stellar particles that become non-disc particles are not found at preferentially large distances, which would be the case if they were being stripped away from the outer regions of the galaxy. This does not mean that stellar stripping does not occur, but we have confirmed that it is not the dominant process in determining the stellar morphological transformation of the satellites. 

It should be noted that stellar particles that change their disciness status (i.e. disc to non-disc) only represent $\sim5-10\%$ of the stellar mass of galaxies at any given snapshot. The majority of the stellar mass of cluster galaxies is comprised of stars that were on non-disc orbits and remain on non-disc orbits in subsequent snapshots ($\gtrsim40$ per cent on average) -- our condition for a galaxy to be discy being set at $\mathit{f}_{\epsilon}>0.4$ -- and disc stars that remain on disc orbits in subsequent snapshots ($\sim10-35$ per cent at each snapshot). 

Additionally, we find that as expected (but not shown for clarity), stars that remain on non-disc configurations between snapshots are on average always the most-centrally concentrated in comparison to all other stellar particles in every satellite. Finally, we do see that stars transitioning from disc-to-non disc stars actually move physically inward on average within their host galaxies.

All these results together indicate that the morphological transformation does not occur simply by stripping away the outer disc mass from the galaxies. Instead, the orbits of the galaxies' star particles are altered under the influence of the cluster, by moving from disc to non-disc orbits. In conclusion, all our findings point to the following interpretation: the morphological transformation of cluster disc galaxies is the result of tidal shocking \citep[e.g.][]{Gnedin99} as the galaxies orbit close to the cluster centre, where tides are strongest and where the quick gravitational perturbations result in heating up disc stars into non-disc orbits. Importantly, however, we argue that these mechanisms in the cluster potential must occur concurrently with the cessation of star-formation, otherwise newly-formed stars would replenish the stellar discs of cluster satellites, as we discuss in more depth in the next section. Finally, the importance of pericentric passages for altering stellar morphology that we find in this analysis is in line with previous findings with controlled experiments whereby more radial satellite orbits appear more effective at inducing morphological changes compared to more circular orbits \citep[e.g.][]{Kazantzidis11}.

\subsection{Morphological transformations of control vs. cluster discs}

\subsubsection{The case of the control galaxies}
Although a larger proportion of the overall population of control discs remain discy to $z=0$, there is a significant fraction, especially at the high-mass end, that does become non-discy (Fig.~\ref{fig:discDestructionFrac}). Pinning down the exact physical mechanisms that are responsible for transforming the stellar morphology of {\it field} galaxies is beyond the scope of this paper. Importantly, the control disc galaxies we have identified for our analysis are meant to mirror the sample selection of satellite galaxies in $10^{14}\MSUN$ clusters, and are not necessarily an unbiased sub-population of central galaxies. Yet, throughout the cluster versus control comparison in the previous sections, we have explored two of the most commonly invoked processes that are thought to induce or mediate the loss of disciness in galaxies -- galaxy-galaxy mergers and feedback from supermassive black holes (Sections \ref{sec:intrinsic}, \ref{sec:mergers} and \ref{sec:agn}).

Following classical results (see Introduction), \citealt{Martin18} used the results of the Horizon-AGN simulation to show that massive spheroidal galaxies (stellar mass $\gtrsim 10^{10.5}\MSUN$) are formed through high mass-ratio mergers ($>$1:10) and that while major mergers appear to be the dominant cause, minor mergers can account for a third of the morphological transformation of galaxies. For intermediate mass galaxies, they also find that while environmental processes like harassment are important, mergers remain a major contributor to the transformation of the total galaxy population (which is in fact dominated by non-satellite galaxies). \citealt{Clauwens18} also find comparable results using the EAGLE simulations, whereby the most massive central galaxies have more spheroidal morphologies that were largely built up through mergers, which lead to the growth of their bulge component without significantly destroying their discs. Similarly, using the original Illustris simulation, \citet{RodriguezGomez17} explored the relationship between galaxy morphology and the ex-situ stellar mass fraction, with the latter being used as a proxy for the overall importance of gas-poor mergers in a galaxy's assembly history. They found that mergers have a dominant effect on the morphology of massive galaxies ($M_{*} \gtrsim 10^{11}\MSUN$), while the spin of the dark matter halo plays a more important role in the morphology of dwarfs ($M_{*} \lesssim 10^{10}\MSUN$). For medium-sized galaxies ($M_{*} \sim 10^{10-11}\MSUN$), morphology was found to depend on a combination of these two factors.

As with these previous results, in this study using the IllustrisTNG simulations, galaxy-galaxy mergers appear to be a fundamental process for the stellar morphological transformation of galaxies that are centrals or are not in high-density environments. In Section~\ref{sec:mergers}, we have seen that field control galaxies that have had at least one major merger since $z=1$ are more likely to become non-discy by $z=0$ than those that have not had any major mergers. Assuming that every galaxy that has had a major merger and is non-discy at $z=0$ was transformed \emph{by the merger event}, major mergers can account for $50\%$ and $76\%$ of the control discs that are non discs at $z=0$ in the mass ranges of $\log{M_{\text{*,acc}}}=9-9.6$ and $10.6-11.6$ respectively, i.e. our lowest and highest mass bin. In the two intermediate mass bins, for TNG50 and TNG100 separately, major mergers can account for $25-40\%$ of the $z=0$ non-disc control sample. When we include the occurrence of minor mergers, these numbers increase to $71\%$ and $92\%$ in the lowest and highest mass bin and $48-67\%$ in the intermediate mass bins. For reference, major and minor mergers since $z=1$ could not account for more than one fifth of the \emph{cluster} discs that are non-discs at $z=0$. However, we do find that not all galaxies that have had mergers since $z=1$ are non-discy at $z=0$, indicating that some discs can survive these mergers or re-form after the event, this likely being dependent on the gas content of the merging galaxies with gas-rich mergers able to reestablish disc morphologies \citep[e.g.][]{Hernquist95}. Note that these numbers do not account for other less dramatic gravitational perturbations, as we have not considered in our analysis very minor mergers, i.e. with mass ratios of $<1/10$, or high-speed encounter or flybys, which could explain the morphological transformation of at least a subset of the simulated galaxies. Yet, we find that in the IllustrisTNG simulations, the frequency of relatively recent major and minor mergers is broadly consistent with the number of control galaxies that transform from discs to non-disc stellar configurations. 

It is beyond the scope of this paper to delineate exactly how galaxy mergers can induce changes in stellar morphologies and we refer the reader to the existing body of literature in this regard, particularly in controlled numerical setups \citep[e.g.][and reference therein]{Hopkins09}. It is also beyond the scope of this paper to disentangle the direction of causality between galaxy mergers and the occurrence, strength or effects of AGN feedback \citep[but see][for a demonstration in IllustrisTNG that BH feedback is required to quench entire populations of galaxies amid the usual merger rates provided by the hierarchical growth of structure]{TNGMethodsWeinberger17, Weinberger18}. However, it is important to point out that, within the IllustrisTNG model, for the more massive galaxies (above a few $10^{10}\MSUN$) feedback from SMBHs does affect galaxies, by halting their star formation and by vacating the gas from their central regions \citep{Truong19,Terrazas20,Davies20,Zinger20}. Furthermore, in Section~\ref{sec:agn}, we have shown that higher than average BH energy injections correlate with larger morphological changes for control galaxies, this effect being somewhat reduced for cluster satellites. In summary, for field galaxies, all our findings point to a combination of both galaxy mergers and AGN feedback as the physical mechanisms responsible for the loss of disciness of field galaxies. Such processes are likely to affect the high-mass control discs more severely than their lower-mass analogues since the more massive galaxies are more likely to have had mergers and also to have experienced larger amounts of energy injections via AGN feedback; these trends explain the increasing fraction of control galaxies that change from discs to non-discs with increasing mass in Fig.~\ref{fig:discDestructionFrac}.

\begin{figure}
    \includegraphics[width=\linewidth]{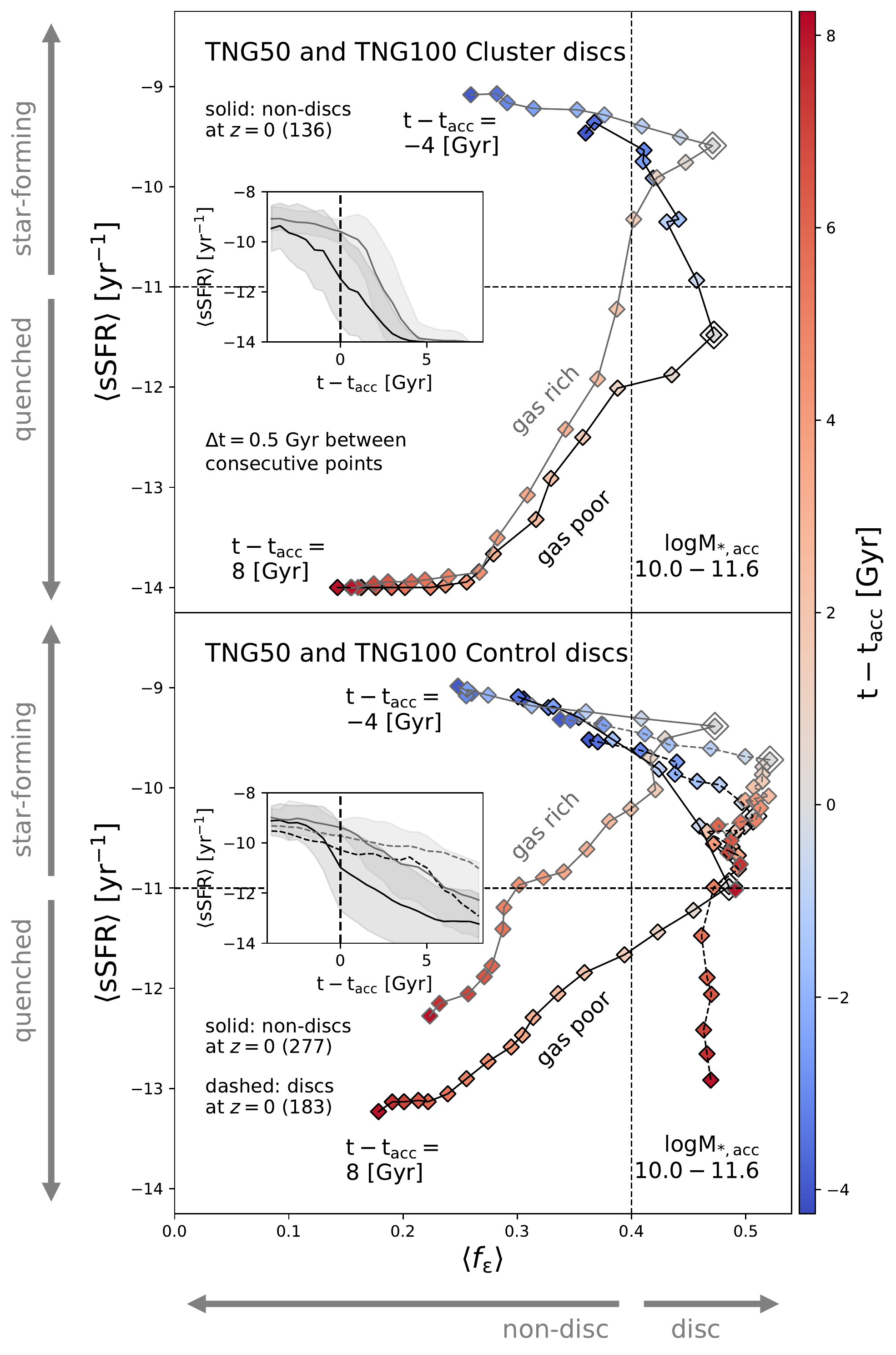}
    \caption{\textbf{Average sSFR versus average circularity fraction $\mathit{f}_{\epsilon}$} for (initially) gas-rich and gas-poor massive galaxies ($M_{\text{*,acc}}>10^{10}\MSUN$). Each point represents the average value of sSFR and $\mathit{f}_{\epsilon}$ for all cluster (top) and control (discs), for all snapshots lying within 0.5 Gyr bins of time relative to accretion. Larger diamond symbols mark the $t-t_{\text{acc}}=0$ data points for each curve. Solid lines show the results for cluster and control discs that are non-discs at $z=0$. Additionally, control discs that remain discs at $z=0$ are shown in the bottom panel; we omit such results for the cluster discs since very few cluster discs remain non-discs at $z=0$. Both TNG50 and TNG100 galaxies are included in the averages. The colours of the datapoints indicate time relative to accretion. Galaxies are classified as gas-rich/poor based on whether their gas fractions at accretion $M_{\text{gas}}/M_{*}(\text{acc})$ are above or below the median for each sample. For the cluster discs, $M_{\text{gas}}/M_{*}(\text{acc})$ ranges from (0-0.2) and (0.2-1) for the gas-poor and gas-rich subsamples respectively; the corresponding ranges for the control discs are (0-0.25) and (0.25-1.35). \emph{Insets:} Average sSFR versus cosmic time for the gas-rich/gas-poor cluster/control discs. Shaded regions show the scatter in 0.5 Gyr bins of time relative to accretion. In the bottom inset panel, we only show the scatter for control discs that become non-discs at $z=0$ (solid lines), but the scatter for those that remain discs (dashed lines) is similar. Initially gas-rich cluster discs continue star-formation for $\sim2$ Gyr after accretion whereas gas-poor ones are quenched almost immediately on average. The trends for the cluster discs that become non-discs are similar, with gas-rich galaxies continuing star-formation for $\sim4.5$ Gyr after `accretion', but gas-poor ones quenching almost immediately. Morphological transformation lags behind quenching for gas-poor cluster and control discs by $\sim1-1.5$ Gyr, but precedes quenching by $\sim0.5$ Gyr for gas-rich cluster discs and $\sim2.5$ Gyr for gas-rich control discs.}
    \label{fig:avgSFREvol}
\end{figure}

\subsubsection{Gas depletion, reduced star-formation, and morphological perturbations} \label{sec:discussionSFRvsFcirc}

The phenomenological comparison between our IllustrisTNG cluster and control samples provides some key insights. There are only two changes in galaxy properties that occur in similar fashion for both the control and cluster galaxies as they transform from discs to non-discs: they see a dramatic reduction of their gas content (Fig.~\ref{fig:changeCircVsChangeMass}, middle column) and of their star-formation rates (Fig.~\ref{fig:changeCircVsChangeDelSsfr}). This is in stark contrast, for example, to what happens to the stellar and DM contents (Fig.~\ref{fig:changeCircVsChangeMass}, left and right panels) and to the galaxy sizes (Fig.~\ref{fig:changeCircVsChangeRhalf}), whereby control (cluster) discs do (do not) keep increasing their DM mass, stellar mass, and stellar sizes as they transform from discs to non-discs. The connection between ``sphericalization'' and star-formation quenching is certainly not novel, but the similarity in behavior that we highlight above and that occurs throughout environments is further corroborated by more subtle dependencies we have discovered.

Firstly, for initial stellar masses of $\gtrsim10^{10}\MSUN$, both cluster and control discs with higher initial (i.e. at accretion) gas fractions exhibit smaller morphological changes and longer transformation timescales than the more gas-poor ones (Section.~\ref{sec:gasFraction}), although the differences are moderate. A similar dependence is found on the \emph{}{rate} of gas reduction, whereby disc galaxies that retain their gas for longer also transform their morphology on longer timescales, with a stronger effect seen for the control discs than in the cluster environment (see Section~\ref{sec:rateGasRemoval}).

One plausible explanation for these connections between morphological changes and gas availability is that larger initial gas reservoirs can allow galaxies to continue forming stars. Since star formation occurs within disc-like, strongly rotationally-supported, thin configurations of gas \citep[as also demonstrated in][for the case of IllustrisTNG galaxies]{TNG50Pillepich19}, a continued replenishment of new stellar material is likely to replenish the stellar discs and to reduce the rate or degree of morphological transformations. Conversely, the removal or eventual consumption of gas within galaxies is necessarily bound to reduce star formation, curtailing the replenishment of stars in disc orbits in the presence of morphological perturbations and therefore enhancing the effects of any mechanism capable of inducing morphological transformation. In fact, it is also plausible to expect that any gravitational perturbation would affect disc stars more strongly in the absence of a ``damping'' that the presence of gas mass could provide. It should be noted that the correlation between morphological change and gas content may simply be the result of an underlying correlation of both properties with stellar mass. To confirm that this is not the case, we have compared $\mathit{f}_{\epsilon}$ and gas mass at $z=0$ in narrow bins of stellar mass: at fixed stellar mass, the lower the remaining gas content of a galaxy, the more non-discy is its final morphology on average, albeit with significant scatter. This result holds for both cluster and control discs, with the cluster (control) discs usually occupying the region of lower (higher) gas mass and $\mathit{f}_{\epsilon}$.

We explore the dependence of the decline of sSFR on initial gas fraction in Fig.~\ref{fig:avgSFREvol}, where for the more massive ($M_{\text{*,acc}}>10^{10}\MSUN$) galaxies in the TNG50 and TNG100 simulations we show average evolutionary tracks on the plane of star-formation rate vs. stellar morphology (i.e. circularity fraction). Average tracks are shown for galaxies with gas fractions above and below the median at accretion (gray vs. black curves, respectively). We have confirmed that although there is significant dependence of gas fraction on stellar mass at accretion, our results remain nearly the same if we were to measure the median gas fraction in bins of stellar mass at accretion. Each datapoint represents the average sSFR and average $\mathit{f}_{\epsilon}$ in 0.5 Gyr bins of \emph{time relative to accretion} as represented by the colour of the datapoint. The solid curves show the results for cluster and control discs that are non-discs at $z=0$. We also show the control discs that remain discs as dashed lines; since there are very few cluster discs that remain discs to present day, we do not show analogous results for the cluster discs. Insets also show average specific star-formation rate histories for cluster (top) and control (bottom) discs separated based on the gas availability at accretion. Here we confirm that galaxies with higher gas content at the time of accretion do indeed have higher sSFRs over somewhat longer periods of time, suggesting that these galaxies continue to form stars over longer timescales. Furthermore, Fig.~\ref{fig:avgSFREvol} demonstrates the close link between star-formation quenching and transformation to non-disc stellar morphology for both cluster and control galaxies, a link that depends not only on environment but also on the available gas content.

We do not aim here to extract quantitative comparisons between the timescales for quenching vs. those for morphological change, as this requires a great amount of precision and consistency across studies in defining the events of accretion, quenching and morphological transformation. We also refer to existing \citep[e.g.][]{Tacchella19} and future analyses that tackle the task of disentangling the timings between the two processes. However, Fig.~\ref{fig:avgSFREvol} does offer a number of insights. Firstly, cluster and control discs that are initially gas-rich are seen to continue star-formation over longer timescales as expected, $\sim2-4.5$ Gyr longer than their initially gas-poor counterparts. In fact, the gas-poor cluster and control discs are quenched essentially at the reference time of `accretion'. Control discs that \emph{remain discs} to $z=0$ also show a similar difference in quenching timescales, with the gas-rich discs being quenched approximately 3 Gyr after their gas-poor analogues. Secondly, for initially gas-poor cluster and control discs that become non-discs at $z=0$, morphological transformation lags behind quenching, occurring on average $1-1.5$ Gyr after accretion (and hence quenching). In contrast, for the gas-rich analogues, morphological transformation precedes quenching by $\sim0.5$ Gyr for the cluster discs, $\sim2.5$ Gyr for the control discs. However, although gas-rich galaxies are seen to transform their morphology \emph{before} they quench, they nonetheless remain discy after accretion for $\sim0.5$ Gyr longer on average than their gas-poor counterparts, consistent with our findings in Fig.~\ref{fig:avgEvolCircIntrinsic}. 

Note that the rapid change in morphology at the reference `accretion' time seen even for the control discs points to the biased nature of the control sample selection, especially in this mass range. Since the control samples were selected to mimic the mass distribution of the cluster satellites, they appear to have found field galaxies that were on the verge of morphological transformation at the `accretion' times. This is even true for the control discs that remain discs at $z=0$, although the change in morphology is milder.

The same analysis for lower mass galaxies ($M_{\text{*,acc}}<10^{10}\MSUN$) yields significantly different results (not shown for brevity), starting with a larger range in the initial gas fractions themselves by a factor of $\sim2$. While the gas-rich cluster discs that are non-discs at $z=0$ quench $\sim3$ Gyr after morphological transformation occurs, analogous control discs do so with a $\sim5$ Gyr delay. Importantly, the key contrast with more massive galaxies is that gas-poor discs also show the same delay in quenching, with the delay being $\sim1.5$ Gyr for the cluster sample and $\sim5$ Gyr for the control sample. Additionally, control discs that \emph{remain} discs at $z=0$ continue to become discier after `accretion', although the gas-poor discs eventually start changing to less discy morphologies. Both gas-poor and gas-rich control discs that remain discs at $z=0$ also remain star-forming and show little difference in their average sSFRs as a function of time relative to accretion.

We conclude by summarizing as follows what we have gathered from this exploration of IllustrisTNG galaxies and previous studies. The necessary conditions for the long-term transformation of galaxy stellar morphologies from discs to non-discs appear to be the concurrent reduction of available gas within galaxies and the availability of mechanisms that are able to perturb pre-existing stars into non-disc orbits. This applies in general terms to both cluster satellites and field galaxies, whereby we argue that the lack of gas seems to be required in order to reduce or halt the formation of new stars that would otherwise replenish the stellar discs (in addition to possibly enhancing the effects of the perturbation mechanisms themselves). However, the physical processes that bring satellite and field galaxies to these conditions can be very different, not just because of the different environment but also depending on galaxy stellar mass. Gas depletion in IllustrisTNG cluster galaxies across masses is achieved chiefly (and quickly, within 2-3 Gyr after accretion) via ram-pressure stripping \citep{Yun19}. This is superimposed on the ejection of (star-forming) gas via BH feedback for galaxies of stellar mass above a few $10^{10}\MSUN$ \citep{Zinger20}, that also acts on massive satellite galaxies. For control (field) galaxies, gas reduction is induced by consumption via star-formation, possibly enhanced by merger events and galaxy interactions \citep[see also][]{Hani20}, in addition to the ejective effects of BH feedback for more massive galaxies. Gas replenishment is unlikely for cluster satellites across the mass spectrum, due to the difficulty of accreting ambient gas during high-velocity orbits as in their high-mass host haloes; it is prevented for massive ($\gtrsim$ a few $10^{10}\MSUN$) field galaxies because the BH feedback in the IllustrisTNG model also offsets the cooling times of galactic gaseous reservoirs \citep{Zinger20}. Finally, the perturbation mechanisms that seem to heat up existing stars from their cold, disc-like orbits appear to be primarily tidal shocking for cluster satellites and galaxy mergers and interactions for field galaxies.

\section{Conclusions \& Summary} \label{sec:conclusion}
We have examined the morphological transformation of disc galaxies within clusters in the TNG50 and TNG100 runs of the IllustrisTNG suite of simulations. Galaxy stellar morphologies have been quantified using a kinematic measure, the circularity fraction $\mathit{f}_{\epsilon}$. We have selected galaxies that are found at $z=0$ within the virial radii of clusters with masses $M_{\text{200c}}\sim 10^{14-14.3}\MSUN$ and $M_{\text{200c}}=10^{14-14.6}\MSUN$ in TNG50 and TNG100, respectively, with corresponding galaxy mass ranges of $M_{\text{*,~z=0}}=10^{8.3-12}\MSUN$ and $M_{\text{*,~z=0}}=10^{9.7-12.5}\MSUN$. Of these, we have considered only those galaxies that are discs at the time of accretion, i.e. with $\mathit{f}_{\epsilon}>0.4$. By following the disc galaxies throughout their evolutionary histories from the time of accretion and comparing their transformation to that of a control sample of centrals mass-matched at the time of accretion, we have been able to isolate the influence of the cluster environment. Our findings lead us to conclude the following:

\begin{itemize}
    \item The cluster environment induces the transformation stellar morphologies of galaxies from discs to non-discs. This effect is relatively more effective than what occurs for control disc galaxies in the field, particularly for galaxies below stellar masses of a few $10^{10}\MSUN$. Nearly all the cluster discs in TNG50 and TNG100 (70-95 per cent, depending on galaxy mass and host assembly history) are non-discy by $z=0$, i.e. they undergo morphological transformation along their orbits within their host clusters. In contrast, about 40-70 per cent of the TNG50 and TNG100 control discs remain discy over the same timescales. The control discs show a decreasing trend in survival fraction of discs with increasing stellar mass, whereas the cluster discs show the opposite trend.\\
    
    \item For both cluster and control discs, more drastic transformations to non-discy morphologies are accompanied by larger gas mass losses and reductions of star formation rates. However, while cluster discs become non-discy accompanied by a loss of DM mass and stagnation in growth or mild loss of stellar mass and stellar sizes, control discs do so along with a considerable growth in stellar and DM mass and stellar sizes.\\
    
    \item Cluster disc galaxies that spend the longest time in the cluster environment (i.e. have the earliest accretion times) experience the largest changes in morphology by $z=0$. Control discs also undergo larger morphological changes given more time, but with significantly more scatter, such that there is a large fraction of the control discs that remains discy even over $\sim8-10$ Gyr timescales.\\
    
    \item Cluster disc galaxies that become non-discs by $z=0$ do so on similar timescales irrespective of their stellar mass at accretion (as long as this is larger than about $10^{9.6}\MSUN$). On the other hand, in the control sample, the most massive galaxies ($M_{\text{*,acc}}\approx10^{10.6-11.6}\MSUN$) evolve to non-discs more rapidly than lower-mass ones. Within our definitions, cluster and massive control disc galaxies transform to non-discy objects in $0.5-4$ Gyr (within $\pm1$-sigma of the galaxy-to-galaxy variations). Importantly, the cluster environment acts to remove any dependence on stellar mass of the timescales to become non-discs that would normally exist in the field.\\
    
    \item In the clusters, galaxies that are found today at closer clustercentric distances and that have had closer and more numerous pericentric passages have transformed to a greater extent than galaxies at larger (pericentric) distances and fewer passages close to the cluster centres. Furthermore, for both cluster and control samples \emph{at the high-mass end}, galaxies with higher initial gas content and which retain their gas over longer timescales also exhibit somewhat smaller changes in stellar morphology and mildly longer timescales of transformation.\\
    
    \item By tracking the radial distributions of disc and non-disc stellar particles in each snapshot that become disc or non-disc particles in the subsequent snapshot, we find that galaxies transform into non-disc morphologies not by the stripping away of disc mass, but rather by the alteration of their stellar particles' orbits, predominantly through tidal shocking at pericentre.\\
    
    \item Control galaxies that have had at least one major merger since $z=1$ are more likely to be non-discy by $z=0$, but this connection does not hold for cluster satellites. The occurrence of a major or minor merger can account for the majority \emph{control} discs that become non-discs at $z=0$.\\
    
    \item For galaxies of initial stellar mass above a few $10^{10}\MSUN$, higher than average energy injection from BH feedback correlates with larger morphological changes, this effect also being present, albeit somewhat weaker, for the massive cluster galaxies. Importantly, we find that BH accretion rates averaged over several galaxies and over long timescales are significantly suppressed for the cluster discs compared to the control discs, at all masses.\\
    
    \item There is a strong connection between star-formation quenching and morphological transformation for cluster and control discs that become non-discs at $z=0$. Gas-poor massive cluster and control discs ($M_{\text{*,acc}}>10^{10}\MSUN$) quench $\sim1-1.5$ Gyr before morphological transformation; for all other galaxies in our samples, quenching occurs after morphological change with the average delay timescale dependent on initial gas content and whether the galaxies are cluster satellites or in the field.
    
\end{itemize}

To our knowledge, this is the first time that the morphological transformation of cluster galaxies has been demonstrated to occur with fully cosmological hydrodynamical models. Due to the combination of high resolution and large volume provided by the TNG50 and TNG100 simulations, we have been able to study the morphological evolution of discs in $10^{14}\MSUN$ clusters in a statistical and unbiased manner. The results show that the cluster environment does impact the morphological evolution of galaxies as expected, resulting in proportionally more non-disc galaxies at $z=0$ than in the central population. However, the impact on massive galaxies is distinct from the impact on lower-mass galaxies and there are important cluster-to-cluster variations due to the diversity of mass assembly histories of the host haloes. 

In the case of massive (initial stellar mass greater than a few $10^{10}\MSUN$) central or field disc galaxies selected at about $z\sim1$, more than 50 per cent of discs are transformed to non-discs by $z=0$ due to a combination of galaxy-galaxy mergers and AGN feedback. The effects of the cluster environment add to such physical mechanisms and may affect the secular processes themselves; for example, we find that cluster discs have on average significantly suppressed BH accretion rates after accretion than their analogues in the field.

Our analyses suggest that, for both cluster satellites and field galaxies, there are two necessary and concurrent conditions for the long-term transformation of galaxy stellar morphologies from discs to non-discs: the long-lasting reduction of available gas within galaxies and the exposure to mechanisms able to remove pre-existing disc stars or perturb them into non-disc orbits. However, the physical processes that bring satellite and field galaxies to these conditions can be very different, not just because of the different environment but also depending on galaxy stellar mass. While for control galaxies gas reduction is induced by consumption via star-formation, possibly enhanced by merger events and galaxy interactions, in addition to the ejective effects of BH feedback above a few $10^{10}\MSUN$, gas depletion in cluster galaxies is achieved chiefly via ram-pressure stripping, in combination with gas ejection via stellar and BH feedback. The perturbation mechanisms that induce the morphological transformation appear to be primarily the heating up of existing stars from their cold, disc-like orbits because of tidal shocking for cluster satellites and galaxy mergers and interactions for field galaxies. However, we argue that a persistent lack of gas is required in order to reduce or halt the formation of new stars that would otherwise replenish the stellar structures with young stars in circular, disc-like orbits. Gas replenishment is arguably unlikely for cluster satellites, due to the difficulty of accreting ambient gas during high-velocity orbits in their high-mass hosts. In massive field galaxies, gas replenishment can be stalled by the same preventive mechanisms that achieve a long-lasting cessation of star formation -- within the IllustrisTNG model, this is AGN feedback and its capability of offseting the cooling times of galactic gaseous reservoirs. Our picture naturally supports the close link between star-formation quenching and transformation to non-disc stellar morphologies across the galaxy mass spectrum and throughout environments.

\section*{Data Availability Statement}
The data that support the findings of this study are available on request from the corresponding author. The data pertaining to the TNG100 simulation are in fact already openly available on the IllustrisTNG website, at \url{www.tng-project.org/data}; those of the TNG50 simulation are expected to be made publicly available within some months from this publication, at the same IllustrisTNG repository.

\section*{Acknowledgements}
The authors would like to thank Andrew Wetzel and Elad Zinger for useful conversations. The IllustrisTNG simulations used in this work have been realized with compute time granted by the Gauss Centre for Supercomputing (GCS): TNG50 under GCS Large-Scale Project GCS-DWAR (2016; PIs Nelson/Pillepich) and TNG100 under GCS-ILLU (2014; PI Springel) on the GCS share of the supercomputer Hazel Hen at the High Performance Computing Center Stuttgart (HLRS). FM acknowledges support through the Program "Rita Levi Montalcini" of the Italian Miur.

%%%%%%%%%%%%%%%%%%%%%%%%%%%%%%%%%%%%%%%%%%%%%%%%%%



\bibliographystyle{mnras}
\bibliography{TNGClusterSpiralMorphology}

%\begin{thebibliography}{99}
%\end{thebibliography}

%%%%%%%%%%%%%%%%%%%%%%%%%%%%%%%%%%%%%%%%%%%%%%%%%%

%%%%%%%%%%%%%%%%% APPENDICES %%%%%%%%%%%%%%%%%%%%%


\appendix

\section{Accretion times} \label{sec:clusterSampleStats}
\label{app:accretionTimes}
\begin{figure}
	\includegraphics[width=\linewidth]{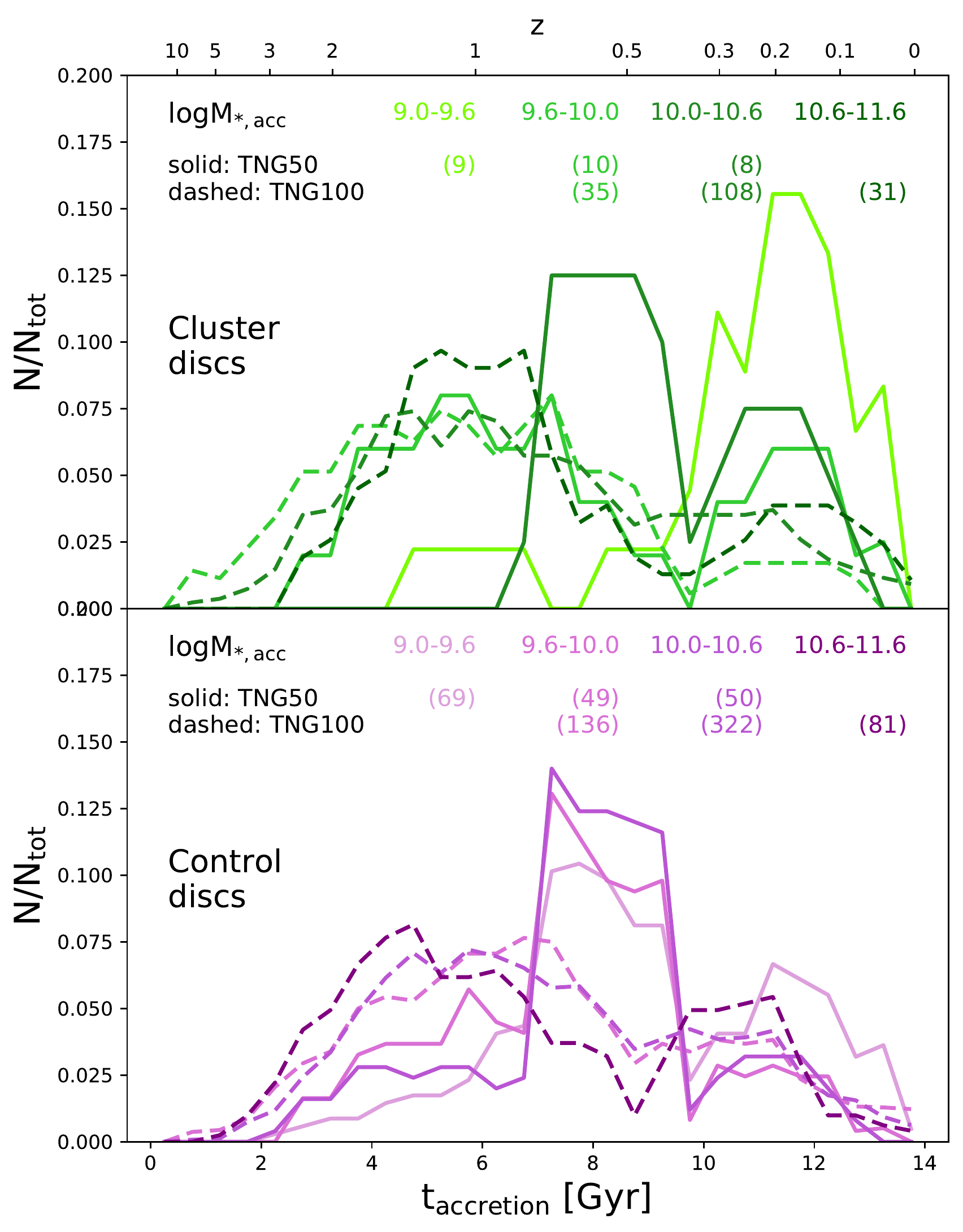}
	\caption{Distribution of accretion times (i.e. the age of the Universe at the time of accretion) of the cluster (top) and control (bottom) discs. `Accretion' for the control samples refers to the accretion time of the cluster galaxy they were mass-matched to. Colours indicate bins of stellar mass at accretion; TNG50 results are shown by solid lines, TNG100 results by dashed lines. The data have been smoothed using a running average in overlapping bins of width 1.25 Gyr after first binning the raw data in bins of 250 Myr in $t_{\text{accretion}}$.} \label{fig:accTimesDist}
\end{figure}

Fig. \ref{fig:accTimesDist} shows the distribution of accretion times of the TNG50 and TNG100 cluster and control samples, separated into bins of stellar mass at accretion. As mentioned in Section \ref{sec:sampleControl}, `accretion' for the control samples refers to the accretion time of the cluster galaxy they were mass-matched to. The data have been smoothed by first binning the raw data into bins of 250 Myr in $t_{\text{accretion}}$ and then calculating a running average in overlapping bins of width 1.25 Gyr. The cluster disc and control disc distributions are not identical, since we mass-match the entire cluster satellite population first and then independently select discs from the cluster and control populations. Additionally, the distributions of the two cluster disc distributions are significantly different due to the merger histories of the host clusters in TNG50 and TNG100 being quite varied. However the Fig. shows that the distribution of accretion times is similar for the cluster and control discs in corresponding mass bins and there is no significant bias in our samples, at least for the TNG100 samples. The distributions for the TNG50 samples are mildly different, partly due to the small sample sizes of the cluster discs.

\begin{figure*}
    \centering
    \includegraphics[width=0.9\linewidth]{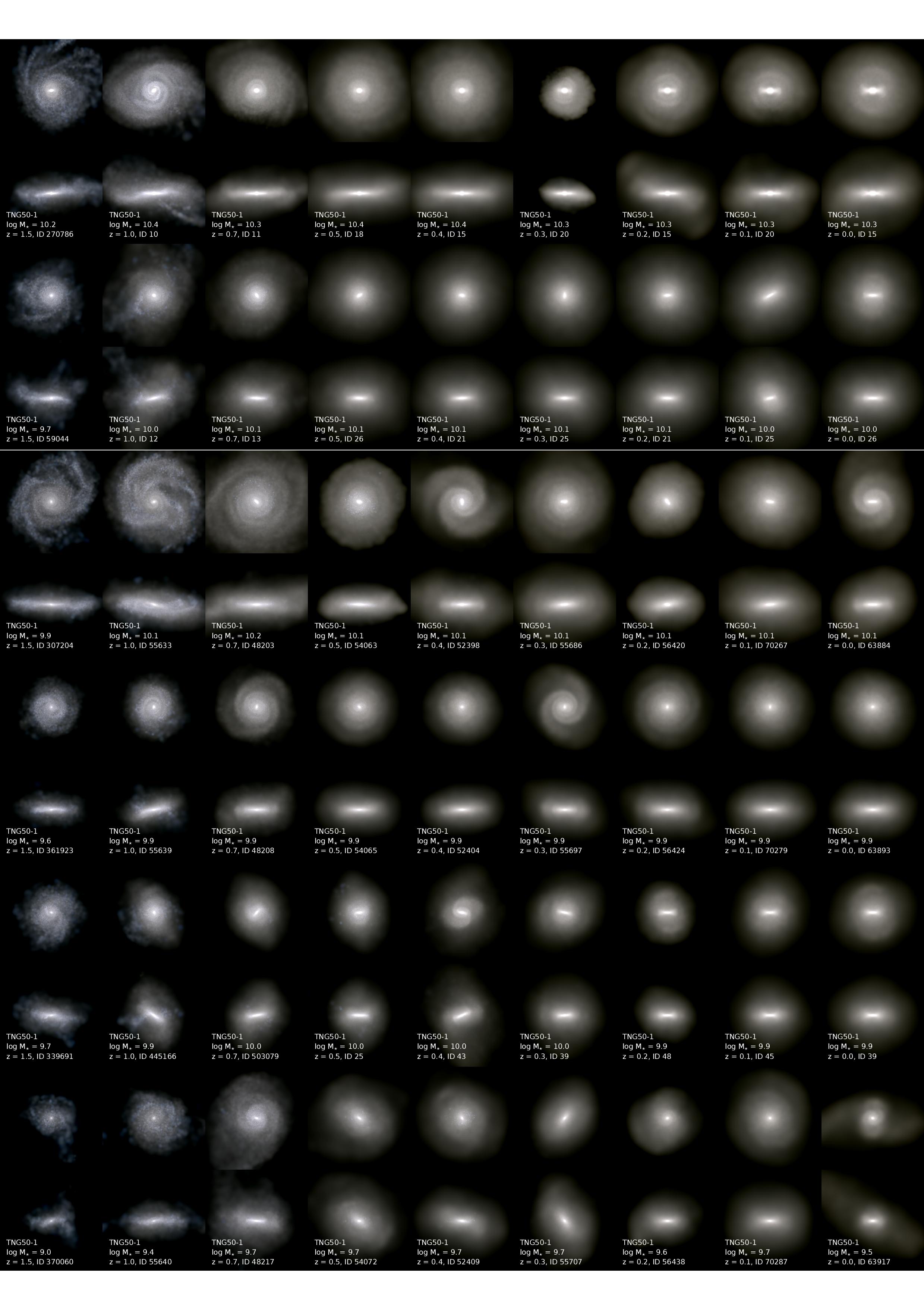}
    \caption{Stamps showing projected stellar composite maps for example TNG50 cluster galaxies with JWST NIRCam f200W, f115W, and F070W filters (rest-frame, no dust), demonstrating their evolution from $z=1.5$ (left) to $z=0$ (right). Odd rows show face-on projections, even rows edge-on projections. Every panel is 40 physical kpc on a side. These satellite galaxies are selected among those that were discy at accretion and become non-discs by $z=0$.} \label{fig:egStampsTNG50}
\end{figure*}

\begin{figure*}
    \centering
    \includegraphics[width=0.9\linewidth]{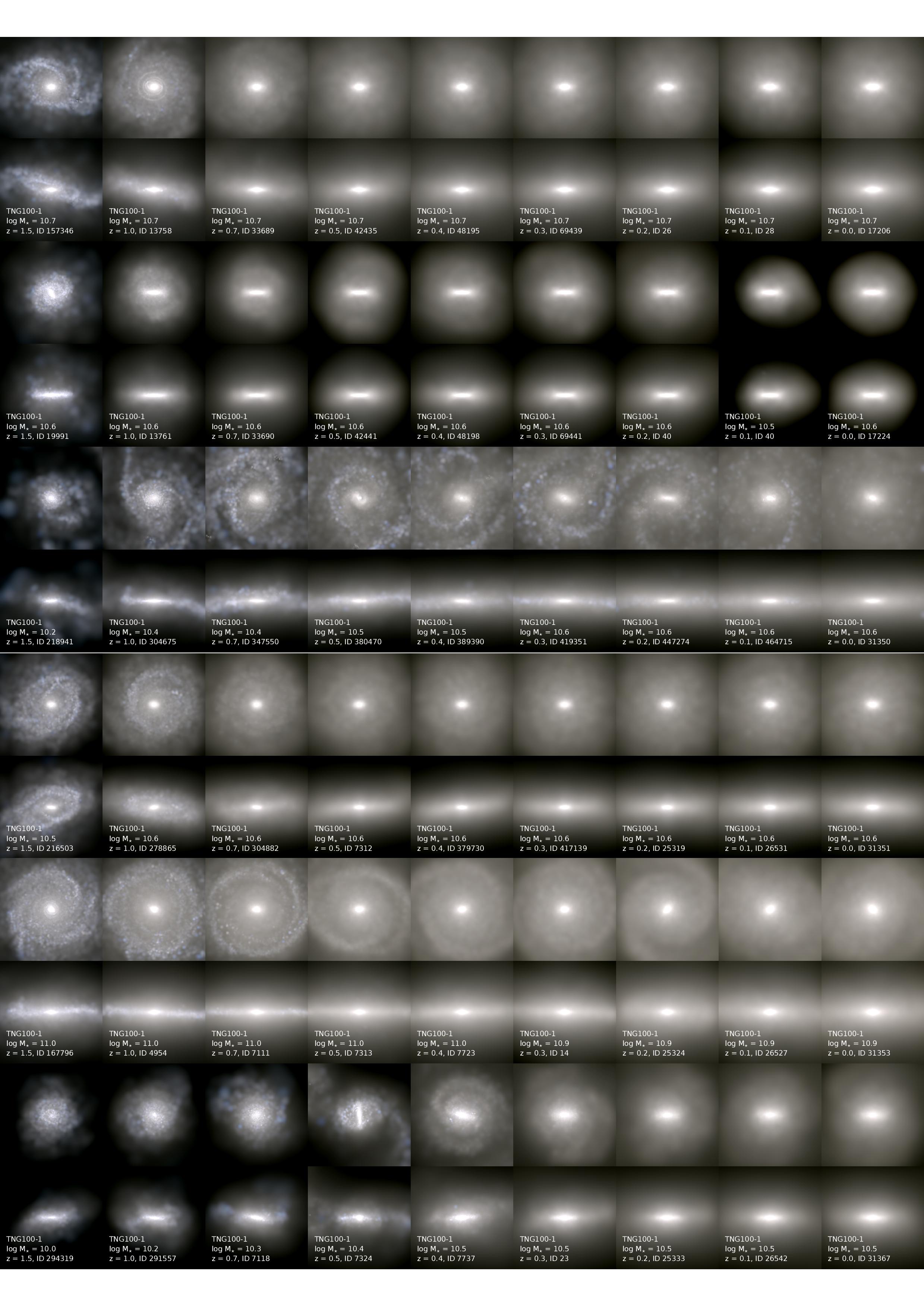}
    \caption{Stamps showing projected stellar composite maps for example TNG100 cluster galaxies with JWST NIRCam f200W, f115W, and F070W filters (rest-frame, no dust), demonstrating their evolution from $z=1.5$ (left) to $z=0$ (right). Odd rows show face-on projections, even rows edge-on projections. Every panel is 40 physical kpc on a side. These satellite galaxies are selected among those that were discy at accretion and become non-discs by $z=0$.} \label{fig:egStampsTNG100}
\end{figure*}

\section{Morphology evolution}
Figs. \ref{fig:egStampsTNG50} and \ref{fig:egStampsTNG100} show projected stellar composite maps for a few example cluster discs from the TNG50 and TNG100 cluster samples (that become non-discs by $z=0$) respectively, demonstrating their evolution from $z=1.5$ to $z=0$. While the cluster discs are initially quite discy and most have distinct spiral arms, at $z=0$, low values of $\mathit{f}_{\epsilon}$ result in a diversity of visual morphologies including bars, tidal features and puffed up discs.

\section{Robustness of morphology measurement at pericentre} \label{app:checksOnSubfind}

Our results have shown that a key factor determining morphological transformation of the cluster discs is their pericentric passage, specifically the pericentric distance of their orbits. These results must be interpreted with caution, as pericentric encounters are precisely where the subhalo-finder \textsc{subfind} would have difficulty separating the subhalo from the cluster central. Therefore, these results could be affected by numerical effects due to the subhalo finding. To test the robustness of the morphology measurements, for cluster discs that are non-discs at $z=0$, we recalculate $\mathit{f}_{\epsilon}$ by including all particles that were gravitationally bound to the galaxy at accretion. If the galaxy continues growing in stellar mass after accretion, we include any added stellar particles until it reaches its peak stellar mass. At each timestep after accretion, we measure a new half mass radius, $r_{\text{*,1/2,ALL AT ACC.}}$, including all such tracked particles and calculate a new circularity fraction, $\mathit{f}_{\epsilon,\text{ALL AT ACC.}}$ within $2\,r_{\text{*,1/2,ALL AT ACC.}}$. By doing so, we can exclude particles that have truly been stripped away from the galaxy, but include those that are still found within it, but are deemed unbound by \textsc{subfind}. By comparing $\mathit{f}_{\epsilon}$ measured using both methods over time for individual galaxies (e.g. in Fig. \ref{fig:CircAtpericentre} for one of the TNG50 galaxies), we have confirmed that although there is some difference between the two morphology measurements, the rapid changes in $\mathit{f}_{\epsilon}$ at pericentre are seen regardless of how we measure the property. Therefore, the morphological changes at pericentre, and indeed at any time along the galaxy's history when it is a satellite, appear to be physical and not artifacts of the subhalo-finding process.

%%%%%%%%%%%%%%%%%%%%%%%%%%%%%%%%%%%%%%%%%%%%%%%%%%

% Don't change these lines
\bsp	% typesetting comment
\label{lastpage}
\end{document}

%% file: host_props_TNG50.tex
CL0 & 1.83 & 1.20 & 3859 & 139	 & 13 & 2 \\
CL1 & 0.94 & 0.96 & 2039 & 87	 & 14 & 6 \\

%% file: host_props_TNG100.tex
CL0 & 3.77 & 1.52 & 2272 & 67	 & 20 & 2 \\
CL1 & 3.81 & 1.53 & 1809 & 61	 & 15 & 2 \\
CL2 & 3.38 & 1.47 & 1527 & 59	 & 26 & 0 \\
CL3 & 1.71 & 1.17 & 1510 & 32	 & 11 & 1 \\
CL4 & 2.54 & 1.33 & 1093 & 54	 & 20 & 0 \\
CL5 & 2.03 & 1.24 & 1178 & 33	 & 11 & 0 \\
CL6 & 2.09 & 1.25 & 960 & 43	 & 15 & 0 \\
CL8 & 2.08 & 1.25 & 867 & 34	 & 15 & 1 \\
CL9 & 2.13 & 1.26 & 986 & 37	 & 8 & 0 \\
CL10 & 1.63 & 1.15 & 780 & 26	 & 8 & 0 \\
CL11 & 1.31 & 1.07 & 760 & 19	 & 5 & 0 \\
CL14 & 1.14 & 1.02 & 472 & 20	 & 4 & 0 \\
CL15 & 1.09 & 1.01 & 449 & 25	 & 7 & 2 \\
CL17 & 1.06 & 1.00 & 421 & 18	 & 9 & 0 \\

%% file: sample_TNG50.tex
All & 226 & 27 & 8 & 678 & 191 & 134 \\
$<9.0$ & 86 & 0 & 0 & 242 & 16 & 8 \\
9.0-9.6 & 66 & 9 & 0 & 193 & 69 & 56 \\
9.6-10.0 & 35 & 10 & 4 & 98 & 49 & 37 \\
10.0-10.6 & 29 & 8 & 4 & 103 & 50 & 29 \\
10.6-11.6 & 10 & 0 & 0 & 42 & 7 & 4 \\

%% file: sample_TNG100.tex
All & 528 & 174 & 8 & 1583 & 542 & 222 \\
$<9.0$ & 1 & 0 & 0 & 3 & 0 & 0 \\
9.0-9.6 & 17 & 0 & 0 & 47 & 3 & 2 \\
9.6-10.0 & 161 & 35 & 1 & 434 & 136 & 70 \\
10.0-10.6 & 232 & 108 & 5 & 691 & 322 & 129 \\
10.6-11.6 & 117 & 31 & 2 & 408 & 81 & 21 \\